\documentclass[twocolumn,pra,superscriptaddress,showpacs,tightenlines]{revtex4}
\usepackage{amssymb}
\usepackage{amsmath}
\usepackage{graphicx}
\usepackage{epsfig}
\usepackage{subfigure}
\usepackage{amsfonts}
\usepackage{txfonts}
\usepackage{CJK}
\usepackage[colorlinks,citecolor=blue, linkcolor=blue,hyperindex,CJKbookmarks,dvipdfm]{hyperref}

\begin{document}

%\begin{CJK*}{GBK}{song}
\title{Quantum statistics of the collective excitations of an atomic ensemble inside a cavity }
\author{Jin-Feng Huang }
\affiliation{Institute of Theoretical Physics, Chinese Academy of
Sciences, Beijing, 100190, China}
\author{Qing Ai }
\affiliation{Institute of Theoretical Physics, Chinese Academy of
Sciences, Beijing, 100190, China}
\author{Yuangang Deng }
\affiliation{Department of Physics, Henan Normal University,
Xinxiang 453007, China}
\author{C. P. Sun }
%\email{suncp@itp.ac.cn}
 \affiliation{Institute of Theoretical
Physics, Chinese Academy of Sciences, Beijing, 100190, China}
\date{\today}
\author{Franco Nori }
\affiliation{Advanced Science Institute, RIKEN, Wako-shi, Saitama,
351-0198 Japan}
\affiliation{Department of Physics, University of Michigan, Ann
Arbor, Michigan 48109-1040 USA}

\begin{abstract}
We study the quantum statistical properties of the collective
excitations of an atomic ensemble inside a high-finesse cavity. In the large-detuning regime, it is
found that the virtual photon
exchange can induce a long-range interaction between atoms, which
results in correlated excitations. In particular, the atomic blockade
phenomenon occurs when the induced long-range interaction effectively
suppresses the double atomic excitation, when the average photon
number takes certain values, which makes the two nearest
energy levels degenerate. We also show that quantum phase
transitions occur in the indirectly-interacting atomic ensemble
when the average photon number reaches several critical points. In
this sense, the quantum statistical properties of the collective
excitations are very sensitive to the change of the average
photon number. Our model exhibits quantum phase transitions similar to the ones in the Lipkin-Meshkov-Glick model. Our proposal could be implemented in a variety of systems including cavity quantum electrodynamics (QED), Bose-Einstein condensates, and circuit QED.
\end{abstract}

\pacs{42.50.Nn, 42.50.Dv, 73.43.Nq} \maketitle

%文章打成单列的命令%
%%%%%%%%%%%%%%%%%%%%%%%%%%%%%%%%%%%%%%%%%%%%%%%%%%%%%%%%%%%%%%%
%\global\columnwidth20.5pc
%\global\hsize\columnwidth\global\linewidth\columnwidth
%\global\displaywidth\columnwidth
%%%%%%%%%%%%%%%%%%%%%%%%%%%%%%%%%%%%%%%%%%%%%%%%%%%%%%%%%%%%%%%%%%

\section{\label{sec:1}Introduction}
In quantum optics, photon statistics reflect the essential
properties of the electromagnetic field~\cite{Glauber1963}. Importantly,
correlated photon counting by the second-order correlation function
can characterize the very quantum nature of light, such as bunching
and antibunching effects~\cite{Walls}, as well as the photon
blockade~\cite{photon blockade1,photon blockade2}, which is also referred to as optical state truncation~\cite{blockade_adam}. The quantum
statistical approach for photon counting~\cite{Lambert2010} is also applicable to other
massive and massless bosons~\cite{Meystre2005}. The collective
excitations of an atomic ensemble could be regarded as an operational
quantum memory~\cite{Luking2001_2003,CPSun2003} and the ensemble behaves as a
boson in the large $N$ limit with low excitations~\cite{GRJin}. Therefore, it
is expected that the quantum statistical approach can also work well
for atomic collective excitations. Moreover, the quantum
correlations of these excitations can also be responsible for double-excitation effects, such as the Rydberg blockade, where double
excitation is strongly suppressed by the dipole-dipole interaction
between highly excited Rydberg atoms~\cite{Rydberg blockade1,Rydberg
blockade2,Rydberg blockade3}.

The atomic blockade is similar to the Coulomb blockade, a
typical mesoscopic phenomenon where a single electron prevents an electric
current from crossing some confined
nanostructure~\cite{coulomb1,coulomb2,coulomb3,Kastner1992_1993}.
Similar blockade effects have been predicted and also observed in
quantum optical system for photons~\cite{photon blockade1,photon
blockade2} and cold atoms~\cite{atom blockade in space,Rydberg
blockade1,Rydberg blockade2,Rydberg blockade3}. Recently, phonon blockade has been studied~\cite{phonon-blockade}. The blockade effect, whereby a single particle prevents the
flow~\cite{coulomb1,coulomb2,coulomb3,photon blockade1,photon
blockade2,atom blockade in space} or excitation of many particles,
provides a mechanism for the precise manipulation of quantum states of
microscopic quantum objects at the level of a single particle. In this
sense, it is essential for the implementation of single-particle-based
quantum devices. The photon blockade effect may have applications in
single-photon sources, needed for the physical implementation
of quantum cryptography protocols~\cite{cryptography}.

In this paper we consider quantum correlation effects for an
atomic system. One of the correlation effects studied is the Rydberg
blockade effect. We consider a similar atomic blockade effect using
an indirect-interaction coupling, which is induced by some confined
photons in a cavity rather than by dipole-dipole interactions
between atoms, as in the Rydberg blockade. Physical properties of atomic ensembles can also be quantified via spin squeezing~\cite{spin-review}.

Specifically, we study the case where an ensemble of two-level atoms are coupled to
a cavity field with a large detuning frequency. The photons in the
cavity can induce excitation hopping among atoms, which form a
collective excited state described by the number of excited atoms. We will consider
the case where the number of excited atoms is similar to the difference between
the numbers of excited atoms and unexcited atoms.
%which is similar to the difference between the numbers of excited
%atoms and unexcited atoms.
Furthermore, the variation of half of this
difference equals the variation of the number of excited atoms.

Similar to the generic Coulomb interaction for the Rydberg
blockade~\cite{Rydberg blockade1,Rydberg blockade2,Rydberg
blockade3}, the induced interaction by cavity photons is also a
long-range interaction and results in inhomogeneous
energy-level spacings. More specifically, the structure of the
energy levels depends on the average photon number. We find that
there will be two degenerate energy levels at an integral multiple
of $1/2$ for the average photon number. If the average photon number slightly deviates from an odd multiple of
$1/2$, these two degenerate levels will become nearly
degenerate but far away from other energy levels. Hence, it is difficult for the atomic
ensemble to transit from the nearly-degenerate levels
to other levels. This shows that the double excitation requires higher
energy, which is off-resonant to two single excitations. Therefore, the atomic blockade effect could occur. If we further change the
average photon number, the pair of nearly-degenerate energy levels shifts far away from
each other, but one of them could end up closer to a neighboring energy
level which was far away from this pair before changing the average photon number. Thus,
the occurrence of atomic blockade can be controlled by the average photon number in the
cavity.

Meanwhile, a quantum phase transition (QPT)~\cite{QPT,XFShi2009,JQYou2010} occurs when
the average photon number is a half-integer, for negative detuning (the difference
between the atomic energy-level spacing and the frequency of the cavity
field). This is partially due to the energy-level crossing under the
above conditions. The ground state changes drastically around the
critical points characterized by the average photon number. This QPT behavior is similar to that of the Lipkin-Meshkov-Glick (LMG) model~\cite{LMG}, which was studied in the quantum-information-process context in, for example, Ref.~\cite{LipkinModel_adam}. In this
sense, we can regard our system as a modification of the
LMG model. However, the critical points in our system are average-photon-number-dependent. This provides a controllable way to manipulate the system
between different phases.

%%%%%%%%%%%%%%%%%%%%%%%%%%%%%%%%%%%%%%%%%%%%%%%%%%%%%%%%%%%%%%%%
\begin{figure}[ptb]
\includegraphics[bb=163 383 431 716, width=8 cm
]{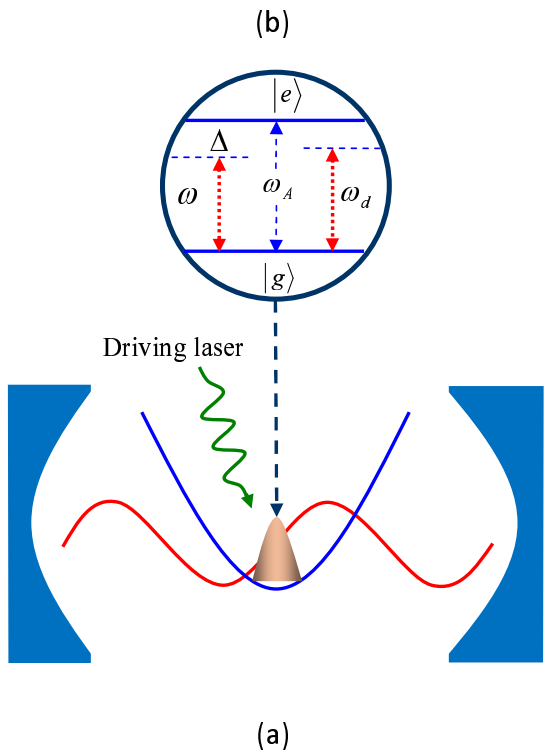} \caption{(Color online) ($a$) Schematic of a cavity
field of frequency $\omega$ coupled to an atomic gas consisting of
$N$ two-level atoms with energy-level spacing $\omega_{A}$. A
driving field of strength $\Omega_{d}$ and frequency $\omega_{d}$ is
applied to the atoms; ($b$) The coupling diagram of one of the
two-level atoms in the cavity. Here, $\Delta$ is the detuning between the atomic level spacing $\omega_{A}$ and the cavity field frequency $\omega$, namely, $\Delta\equiv\omega_{A}-\omega$, and $\omega_{d}$ is the frequency of the drive.} \label{fig1}
\end{figure}
%%%%%%%%%%%%%%%%%%%%%%%%%%%%%%%%%%%%%%%%%%%%%%%%%%%%%%%%%%%%%%%%%

To characterize various correlation phenomena of the atomic
collective excitation, such as the atomic blockade and sensitivity of
the QPT~\cite{QPT sensitivity1, QPT sensitivity2, JFHuang,
QAi,Emary2003}, we introduce a generalized second-order coherence function by replacing the annihilation (creation) $a$ ($a^{\dag}$) operator
of photons in the usual second-order coherence function of photons with the
lowering (raising) $ J_{-}$ ($J_{+}$) operator of the collective atomic
excitations. We prove that the antibunching effect occurs near odd
multiples of $1/2$ for the photon number, which implies that the double atomic excitation is suppressed. We also find significantly different
behaviors on either side of the critical points.

This paper is organized as follows. In Sec.~II, we describe the
system based on the Dicke model~\cite{Dicke,JFHuang}. The
effective Hamiltonian is given in terms of the collective excitation
of the atomic ensemble, and the ground state is analyzed for
different APNs. In Sec.~III we then coherently drive the atomic
ensemble and derive the effective Hamiltonian near two critical
points $n^{c}_{a}=1/2$ and $n^{c}_{a}=j-1/2$. In
Sec.~IV, we introduce the generalized second-order coherence function and calculate the statistical properties of the excitations of
the atomic ensemble in the cases with and without dissipation. We discuss the atomic blockade effect and sensitivity
of the QPT to the photon number in Secs.~V and~VI, respectively.
Finally, we present our conclusions in Sec. VII. The explicit form of the
parameters used in Secs.~IV and~VI are given in the appendix.

%In the following Sec. IV we plot some figures to explicitly show the
%atomic excitation blockade effect. In Sec. V, we numerically
%addresses the generalized second-order coherence function regulated
%by photons number and plot some figures which shows explicitly the
%critical behavior. We numerically consider the dissipation case for
%the generalized second-order coherence function in Sec. VI. Finally,
%we draw our conclusion in Sec. VII. The detailed coefficients in
%Sec. III are given in the Appendix.

\section{\label{sec:2}Quantum criticality of an atomic ensemble strongly
coupled to a cavity field}

\subsection{Model and Hamiltonian}

As shown in Fig.~\ref{fig1}, the system we consider consists of an ensemble of
atoms confined to a single-mode cavity of frequency $\omega$. The
cavity field is described by the annihilation (creation) operator
$a$ ($a^{\dag}$). This model can be implemented in a variety of systems including cavity QED~\cite{Haroche1983}, Bose-Einstein condensates~\cite{Esslinger 2007}, and circuit QED~\cite{Twamley2010}.

%This setup is feasible in the present
%experiments~\cite{Esslinger 2007,Twamley2010}. This current
%technique of trapping ultra cold atoms in cavity has also been
%realized experimentally for Bose-Einstein condensate
%recently~\cite{Esslinger 2007}.

Our model is described by the Dicke
Hamiltonian~\cite{JFHuang,Dicke,Yong Li2006,Yong Li2010,Lambert2009,Carmichael2007,Domokos2010,Twamley2010}
(hereafter, we take $\hbar=1$),
\begin{equation}
H_{1}=\omega
a^{\dag}a+\frac{\omega_{A}}{2}\sum_{\ell=1}^{N}\sigma_{z}^{\left(
\ell\right) }+\frac{g_{0}}{\sqrt{N}}\sum_{\ell=1}^{N}\left( a^{\dag}\sigma
_{-}^{\left( \ell\right) }+a\sigma_{+}^{\left( \ell\right) }\right)
\text{,}     \label{DickeH}
\end{equation}
under the rotating-wave approximation. Here, we use the Pauli matrices $\sigma_{z}^{(j)}=|e%
\rangle_{jj}\langle e|-|g\rangle _{jj}\langle g|$, $\sigma_{+}^{(j)}=|e%
\rangle_{jj}\langle g|$, and $\sigma _{-}^{(j)}=|g\rangle_{jj}\langle e|$ to
describe the atomic transition of the $j$th atom with energy-level spacing $%
\omega_{A}$, where $\left\vert e\right\rangle _{j}$ and $\left\vert
g\right\rangle _{j}$ are the excited and ground states of the $j$th
atom, respectively.

  For an atomic gas with size smaller than the wavelength~\cite
{JFHuang,Emary2003}, we assume that all the atoms are located near
the origin and interact with the cavity field at the
homogeneous coupling rate $g_{0}/\sqrt{N}$. Here, the factor
$\sqrt{N}$ in the denominator of the coupling strength originates from
the fact that the coupling strength is inversely proportional to the
square root of the volume of the cavity field $ 1/\sqrt{V}$. The
volume $V$ is approximately equal to the total volume occupied by the
atoms, which is $N$ times the volume of a single atom. Hence we can write the
factor $\sqrt{N}$ explicitly in the coupling strength.

We would like to point out that, the superradiant phase transition based on the Dicke model in a real atomic system does not exist due to the inclusion of electromagnetic vector potential $A^{2}$ term~\cite{JFHuang,Yong Li2010,Zacowick1975,Rzazewski1979}. However, the following arguments about QPT are based on the LMG model~\cite{LMG}, which will be derived from the above Dicke model, even including the $A^{2}$ term. The similar $A^{2}$ term ($V^2$ term) in circuit QED system will not influence the Hamiltonian significantly, except for just a little shift of the critical point~\cite{Ciuti2010}.

The atoms we consider are largely detuned from the frequency
$\omega$ of the cavity field; namely, the detuning $\Delta$
($\equiv\omega_{A}-\omega$) is much larger than the corresponding
coupling strength $g_{0}/\sqrt{N}$, that is, $\left\vert
\Delta\right\vert $ $\gg$ $\left\vert g_{0}/\sqrt{N}\right\vert $.
In this case, one can use the Fr\"{o}hlich-Nakajima
transformation~\cite{Frohlich,Nakajama} (or adiabatic elimination
method), to obtain the effective Hamiltonian,
\begin{eqnarray}
H_{1} & =&\omega a^{\dag}a+\frac{1}{2}\left( \omega_{A}+W\right) \sum
_{\ell=1}^{N}\sigma_{z}^{\left( \ell\right) }+W\sum_{\ell=1}^{N} a^{\dag}a\sigma_{z}^{\left(
\ell\right) }  \notag \\
& &+\frac{1}{2}W\sum_{\ell_{1},\ell_{2}=1}^{N}\left( \sigma_{+}^{\left( \ell_{1}\right)
}\sigma_{-}^{\left( \ell_{2}\right) }+\sigma_{-}^{\left( \ell_{1}\right)
}\sigma_{+}^{\left( \ell_{2}\right) }\right),   \label{H2}
\end{eqnarray}
where $W=g_{0}^{2}/(N\Delta)$
is the strength of the effective interaction among the atoms, which
is induced by the virtual photon exchanges. The form of the Hamiltonian is very similar to
the dipole-dipole interaction of atoms in free space. We note
that the Fr\"{o}hlich-Nakajima transformation is equivalent to the
approach based on the adiabatical elimination and some perturbation
theories \cite{Sun}. Furthermore, the photon number $a^{\dag}a$ becomes a conserved number.

\subsection{Symmetric Hilbert Space and the LMG model}

We now describe the Hilbert space of the symmetric excitation. The
Hilbert space of $N$ two-level atoms is spanned by $2^{N}$ basis
vectors $\{|g_{l}\rangle, |e_{l}\rangle\}$ with $l=1,2,\cdots,N$. In
the present case, all the atoms have identical transition frequencies and
coupling constants with the cavity field. Here, we consider the symmetric collective excitation subspace $V^{[j]}$ of dimension $(N+1)$.
%The Hilbert space is reduced into a
%subspace $V^{[j]}$ of $(2N+1)$ dimension to support the symmetric
%collective excitation.
We now introduce the collective operators,
\begin{equation}
J_{\pm}=\sum_{\ell=1}^{N}\sigma_{\pm}^{\left( \ell\right) },\hspace{0.5cm}J_{z}=%
\frac{1}{2}\sum_{\ell=1}^{N}\sigma_{z}^{\left( \ell\right) },   \label{J}
\end{equation}
which obey the following angular momentum commutation relations,
\begin{equation}
\lbrack J_{z},J_{\pm}]=\pm
J_{\pm},\hspace{0.5cm}[J_{+},J_{-}]=2J_{z}.
\end{equation}
Furthermore, we define the Dicke basis vectors $|j,m\rangle$ ($j=N/2$,
$m=-j,-j+1,\cdots,j-1,j$), which satisfy
$J^{2}|j,m\rangle=j(j+1)|j,m\rangle$, and
$J_{z}|j,m\rangle=m|j,m\rangle$. One can conclude straightforwardly from Eq.~(\ref{J}) that the magnetic quantum number $m$ equals the half difference between the numbers of excited atoms and the ground state atoms.
In terms of the Dicke states, the symmetric excitation subspace, $V^{[j]}$, is
\begin{eqnarray}
|j,m\rangle \:=\:\mathcal{N}_{m}J^{j+m}_{+}|j,-j\rangle
\:=\:\mathcal{N}_{m}\left[\sum_{\ell=1}^{N}\sigma_{+}^{(\ell)}\right]^{j+m}|G\rangle,\label{state_jm}
\end{eqnarray}
where $\mathcal{N}_{m}=\sqrt{(j-m)!/[(2j)!(j+m)!]}$ and
$|G\rangle=|g_{1},g_{2},\cdots,g_{N}\rangle$.

 %The atomic collective operator $J_{z}$ denotes the collective
%population of the atoms and $J_{\pm}$ represents the symmetric
%collective transitions, alternatively speaking, $J_{\pm}$ denotes
%the change of the number of excited atoms. We notice that the
%excitation created by $J_{+}$ is collective, namely, the obtained
%state by acting $J_{+}|j,m=-j\rangle =\mathcal{N}_{-j+1}\left[\sum_{j=1}^{N}\sigma_{+}^{(j)}\right]|g_{1},g_{2},\cdots,g_{N}\rangle$.
%It follows from Eqs.~(\ref{J}) and~(\ref{state_jm}) that the magnetic quantum
%number $m$ equals the half of the difference between the excited atom number and the unexcited atom number.
According to Eq. (\ref{J}), we can find
\begin{eqnarray}
J_{\pm}|j,m\rangle&=&\sum^{N}_{\ell=1}\sigma^{(\ell)}_{\pm}\:|j,m\rangle  \nonumber\\ %\label{jm}
&=&\sqrt{(j\pm m+1)(j\mp m)}\:|j,m\pm 1\rangle. \label{jmm}
\end{eqnarray}
It follows from Eq.~(\ref{jmm}) that, the ladder operators $J_{\pm}$ describe the action of pumping one more ($J_{+}$) or less ($J_{-}$) atom from
the ground state $\vert g\rangle $\ to the excited state $\vert
e\rangle $. Accordingly, the magnetic quantum number $m$ increases or
decreases by one. %, as shown in Eq.~(\ref{jmm}).
Therefore, when the ladder
operator $J_{+}$ acts on the collective-excitation state $s$ ($0\leq s\leq N$) times,
there will be $s$ atoms being excited, and the magnetic quantum number $m$ will
increase by $s$ accordingly: namely $|j,m\rangle\rightarrow |j,m+s\rangle$, which is
implied in Eq.~(\ref{state_jm}). As for the ladder operator $J_{-}$, the effect is inverse.
%And if it increases
%(decreases) by one, the number of excited atoms is also increases
%(decreases) by one correspondingly. This means there is one more
%(less) atom excited from the ground state $\left\vert g\right\rangle
%$\ to the excited state $\left\vert e\right\rangle $. Hence the
%action of transition operator $J_{+}$ ($J_{-}$) is to pump one more (less) atom from
%the ground state $\vert g\rangle $\ to the excited state $\vert
%e\rangle $. %Conversely, $J_{-}$, the hermitian operator of $J_{+}$,
%represents the process of pumping one more atom from the excited
%state $\vert e\rangle $\ to the ground state $\vert g\rangle $.
Therefore, the variance of the magnetic quantum number $m$ represents the
variance of the atomic-collective-excitation number.

In terms of the above collective operators, the Hamiltonian (\ref%
{H2}) can be rewritten as
\begin{equation}
H_{1}=\omega a^{\dag}a+\left( \omega_{A}+W\right) J_{z}+2Wa^{\dag}aJ_{z}+%
\frac{W}{2}\left( J_{+}J_{-}+J_{-}J_{+}\right).
\end{equation}
In the interaction picture defined with respect to the free
Hamiltonian,  $H_{\rm free}=\omega a^{\dag}a+(\omega_{A}+W)J_{z}$, the Hamiltonian reads%
\begin{equation}
H_{1}^{\left( I\right) }=\varepsilon\left( \hat{n}_{a}\right) J_{z}+\frac{W}{%
2}\left( J_{+}J_{-}+J_{-}J_{+}\right),   \label{H1}
\end{equation}
where $\hat{n}_{a}=a^{\dag}a$ and $\varepsilon\left( \hat{n}%
_{a}\right) =2W\hat{n}_{a}$.
The effective Hamiltonian~(\ref{H1}) is photon-number dependent.
This is a special case of the LMG model~\cite {LMG} with $V=0$. The LMG model can also be implemented using superconducting circuits~\cite {Tsomokos2008,Larson}.
Through the relations $(J_{+}J_{-}+J_{-}J_{+})/ {2}
=J_{x}^{2}+J_{y}^{2}$, the Hamiltonian can be expressed as
\begin{equation}
H_{1}^{\left(I\right) }=-W\left[ \left( J_{z}-\hat{n}_{a}\right)
^{2}-\hat{n }_{a}^{2}-J^{2}\right].   \label{H3}
\end{equation}
As is well known, the LMG model possesses a critical point, at which a
QPT occurs. On either side of the critical point, the number of
excited atoms of the ground states are different; thus the ground
states are essentially different~\cite{LMG QPT1,LMG QPT2,QPT
sensitivity2}. In our system, a similar critical point also
exists. To see this effect explicitly, we calculate the ground
state for the above Hamiltonian in the next section.

The last two terms of Eq.~(\ref{H1}) describe the
interaction among atoms induced by photons in the cavity. This
interaction between atoms is intrinsically caused by the hopping of
photons between different atoms. And the hopping of
photons induces a second-order indirect interaction among atoms. On account of this interaction, the
system shows an obvious nonlinearity with respect to the excitation
number, as shown by Eq.~(\ref{H3}).

\subsection{Quantum Phase Transition Behavior of the Ground state}

We now analyze the discontinuous change of the ground state symmetry when
varying the photon number. For a given Fock state of the field,
$\varepsilon( \hat{n}_{a}) $ is a definite $c$ number. For a general
photon state $|\psi\rangle$ we replace $\varepsilon(\hat{n}_{a})$ by its mean value
such as $\varepsilon\left( \left\langle \hat{n}_{a}\right\rangle
\right) $ [or $ \varepsilon\left( n_{a}\right)$] when our studies
only concern the atomic ensemble. According to Eq.~(\ref{H3}), the eigenstates of the
system are the common eigenstates of $\{J^{2},$ $J_{z}\}$: $\left\{ \text{ }%
\left\vert j,m\right\rangle ;m=-j,-j+1,\ldots,j-1,j\right\} $, for
$j={N}/{2} $, that is,
\begin{equation}
H_{1}^{\left( I\right) }\left\vert j,m\right\rangle |\psi\rangle=E_{m}^{\left( 0\right)
}\left\vert j,m\right\rangle|\psi\rangle,
\end{equation}
with eigenenergies,
\begin{equation}
E_{m}^{\left( 0\right) }=-W\left[ \left( m-n_{a}\right)
^{2}-n_{a}^{2}-j\left( j+1\right) \right] \equiv\omega_{m}\text{.}
\end{equation}
Clearly, the ground state is photon-number dependent, that is,%
\begin{equation}
\left\vert G\right\rangle =\left\{
\begin{array}{c}
\left\vert j,[n_{a}]\right\rangle \text{, \ \ \ \ \ }0\leqslant n_{a}\leq j-%
\frac{1}{2}\text{,\ \  \ \  }\Delta<0, \\
\left\vert j,j\right\rangle \text{, \ \ \ \ \ \ \ \ \ \ \ \ \ \ }n_{a}\geq j-%
\frac{1}{2}\text{,\  \  \ \ }\Delta<0, \\
\left\vert j,-j\right\rangle \text{ or }\left\vert j,j\right\rangle \text{,
\ \ \ \ \ \ }n_{a}=0\text{, \ \ \ \ \ }\Delta>0, \\
\left\vert j,-j\right\rangle \text{, \ \ \ \ \ \ \ \ \ \ \ \ \ }n_{a}>0\text{%
, \ \ \ \ \ }\Delta>0,%
\end{array}
\right.
\end{equation}
where $[n]$ denotes the (half) integer nearest to $n$. This fact
means that the \textit{ground state symmetry changes suddenly when the photon number
is varied} from one domain to another.
%%%%%%%%%%%%%%%%%%%%%%%%%%%%%%%%%%%%%%%%%%%%%%%%%%%%%%%%
%\begin{figure}[!htb]
%%\enewcommand{captionlabeldelim}{}
%%\centering
%\label{fig2:a}
%\psfig{file=ground_state4_a.eps,bb=79 192 532 727,width=8 cm,origin=br,angle=0}  \vspace{0.5in}
%\label{fig2:b}
%\psfig{file=ground_state4_b.eps,bb=67 310 526 640,width=8 cm,origin=br,angle=0}
% %% label for entire figure
%\caption{(Color online) Diagram of the ground state of atoms
%consisting of $ N $ two-level atoms controlled by the cavity photon
%number when $\Delta<0$. ($a$) Diagram of the energy levels
%versus magnetic quantum number $m$. The upper figure in the above:
%the ground state locates at $m=n_{a}=j-(n+1)/2$; The lower figure in
%the above: the two degenerate ground states locates at
%$m=j-(n+1)/2$, $j-(n-1)/2$, respectively, while $n_{a}=j-n/2$; ($b$)
%Diagram of the ground states corresponding to different average photon number in the
%cavity.}\label{fig2}
%\end{figure}
%%%%%%%%%%%%%%%%%%%%%%%%%%%%%%%%%%%%%%%%%%%%%%%%%%%%%%%%
%%%%%%%%%%%%%%%%%%%%%%%%%%%%%%%%%%%%%%%%%%%%%%%%%%%%
\begin{figure}[ptb]
\begin{center}
\includegraphics[bb=80 83 519 767, width=8.3 cm]{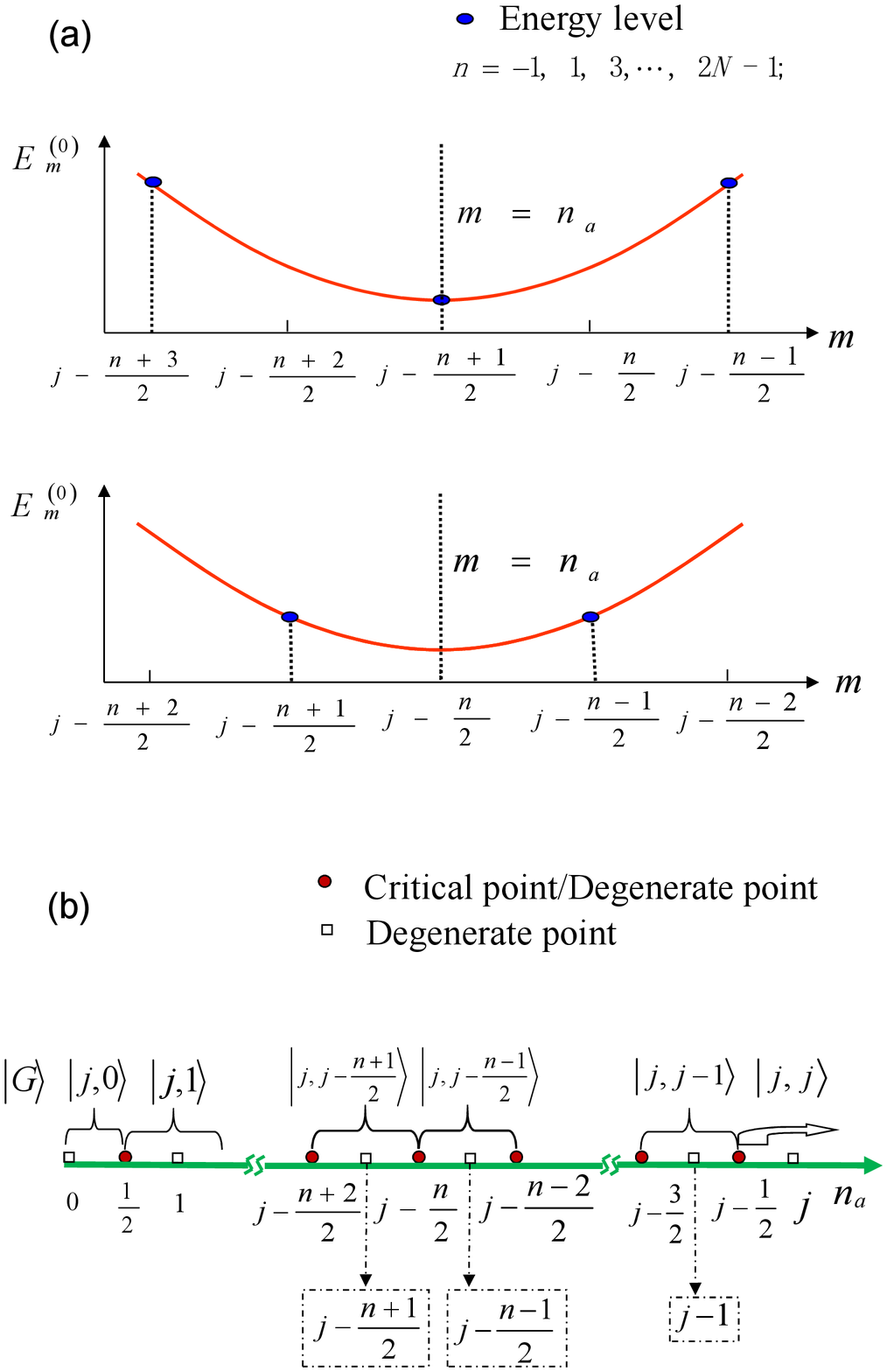}
\caption{(Color online) Diagram of the ground state of atoms
consisting of $ N $ two-level atoms controlled by the cavity photon
number when $\Delta<0$. ($a$) Diagram of the energy levels
versus the magnetic quantum number $m$. The upper figure in ($a$) shows:
the ground state located at $m=n_{a}=j-(n+1)/2$; The lower figure in
($a$) shows: the two degenerate ground states located at
$m=j-(n+1)/2$, $j-(n-1)/2$, respectively, while $n_{a}=j-n/2$; ($b$)
Diagram of the ground states corresponding to different average photon numbers in the
cavity.}\label{fig2}
\end{center}
\end{figure}
%%%%%%%%%%%%%%%%%%%%%%%%%%%%%%%%%%%%%%%%%%%%%%%%%%%%%%%

In the following discussions, we restrict the analysis to the
negative detuning $ \Delta <0$. As shown in Fig.~\ref{fig2}, \textit{when the value of the photon number $n_{a}$ is varied} in the domain of
$0\leqslant n_{a}\leqslant j-1/2$, the atoms will experience
\textit{different ground states, which implies that QPTs occur}.

There are
energy-level crossings at $n_{a}^{c}=j-n/2$, ($n=1,$ $3,$ $5,$
$\cdots,$ $2j-1$). In the domain $j-(n+2)/2<n_{a}<j-{n}/{2}\,$, the
ground state of the system is $\left\vert G\right\rangle =\left\vert
j,j-(n+1)/2\right\rangle \,$, where as in the next domain
$j-n/2<n_{a}<j-(n-2)/2$, the ground state of the system is $\vert
G\rangle =\vert j,j-(n-1)/2\rangle$. If $n_{a}$ increases from
$j-(n+2)/2$ to $j-(n-2)/2$, the energy level of the excited state
crosses the energy level of the ground state at $n_{a}^{c}=j-n/2$. At
the level crossing, the excited state $\vert j,j-(n-1)/2\rangle$ and
the ground state $\vert j,j-(n+1)/2\rangle $ are degenerate. On the
right side of this critical point $n_{a}^{c}$, the original excited
state $ \vert j,j-(n-1)/2\rangle $ in the domain of
$j-(n+2)/2<n_{a}<j-n/2$ will become a new ground state for the
system in the domain of $j-n/2 <n_{a}<j-(n-2)/2$, which implies that a QPT
occurs. In this sense, we can use the average photon number $n_{a}$ to
control the occurrence of the quantum phase transition. At the critical point $
n_{a}^{c}=j-n/2$, both $\vert j,j-(n+1)/2\rangle $ and $\vert
j,j-(n-1)/2\rangle $ are the ground states. Moreover, at this
point, the ground state is highly degenerate, thus the system is in a
symmetric phase.

In other domains, namely, when $\Delta<0$ and
$n_{a}>j-1/2$, or, $\Delta>0$, the ground state is $\vert j,j\rangle $ or $\vert j,-j\rangle $.
In these cases, all the atoms are fully polarized. As all the
two-level atoms can be considered as quasispins, the system is
ferromagnetic in this domain, and the rotational symmetry is broken.
Thus the system is in a symmetry-broken phase. Notice that in the left vicinity of the critical points $n_{a}^{c}$, under the
condition $\Delta<0$, the ground state is $ \vert j,m=[n_{a}]\rangle
$ and possesses one less atomic excitation than that in the first excited
state $\vert j,[n_{a}]+1\rangle $. It is clear that $ \vert
j,[n_{a}]\rangle $ and $\vert j,[n_{a}]+1\rangle $ are nearly
degenerate, but their energies are much less than that of $\vert
j,[n_{a}]+2\rangle $. Thus, there exists an energy gap that makes
exciting two more atoms difficult, but easy for exciting one more
atom. We call this effect ``atomic blockade."

\section{Driven atomic ensemble}

As there exists a level crossing for the photon-dressed atomic
ensemble at $ n_{a}=n_{a}^{c}$, we apply a weak classical driving
to the atomic ensemble. The interaction is described by the Hamiltonian,
\begin{equation}
H_{2}=\Omega\sum_{\ell=1}^{N}\left( \sigma_{-}^{\left( \ell\right)
}e^{i\omega_{d}t}+\sigma_{+}^{\left( \ell\right) }e^{-i\omega_{d}t}\right),
\end{equation}
where $\Omega$ is the Rabi frequency and $\omega_{d}$ is the driving
frequency of the drive. The total Hamiltonian $H=H_{1}+H_{2}$ becomes
\begin{equation}
H^{(R)}=H_{1}^{\left( I\right) }+\left( \omega_{A}+W-\omega_{d}\right)
J_{z}+\Omega\left( J_{-}+J_{+}\right)
\end{equation}
in a rotating frame with rotation
$\exp[i(\omega_{d}J_{z}+\omega a^{\dag}a)t]$. In this driven case, the photon number $a^{\dag}a$ still is a conserved number. Therefore the photon number does not change in the dynamical evolution even though we apply a classical driving field. As a result, we can treat the photon number as an independent external parameter, which is decoupled from the atomic dynamics. We tune the frequency
$\omega_{d}$ to satisfy the resonance condition $\omega
_{A}+W-\omega_{d}=0$. Then the simplified Hamiltonian is $
H^{(R)}=H_{1}^{\left( I\right) }+H^{\prime}$ with $H^{\prime
}=\Omega\left( J_{-}+J_{+}\right) $. When the optical field is
prepared in a coherent state $\vert\alpha\rangle$, the Hamiltonian,
after this average $\hat{n}_{a}\rightarrow{n}_{a}=\langle
\hat{n}_{a}\rangle$, reads
\begin{equation}
H^{(R)}=-W\left[ \left( J_{z}-n_{a}\right) ^{2}-n_{a}^{2}-J^{2}\right]
+\Omega( J_{-}+J_{+}),
\end{equation}
where $ \langle \hat{n}_{a}\rangle=|\alpha|^{2}$, for $n_{a}\equiv1/2+\delta$. Here $\delta$ is
the deviation from the degenerate (critical) point. To see if
the atomic blockade effect occurs, we express the above averaged
Hamiltonian in the angular momentum basis as
\begin{eqnarray}
H^{(R)}&=&\sum_{m=-j}^{j}\omega_{m}\vert j,m\rangle \langle j,m\vert \nonumber \\
&&+\sum_{m=-j}^{j-1}\Omega_{m+1}\left( \left\vert
j,m+1\right\rangle \left\langle j,m\right\vert +\text{h.c.}\right),
\label{H}
\end{eqnarray}
where $\Omega_{m}=\Omega\sqrt{\left( j-m+1\right) \left( j+m\right)
}$. We can then more readily observe the transition from
$|j,m\rangle$ to $ |j,m+2\rangle$ by exciting two more atoms around
the critical point $n_{a}^{c}$.
%%%%%%%%%%%%%%%%%%%%%%%%%%%%%%%%%%%%%%%%%%%%%%%%%%%%%%
%\begin{center}
%\begin{figure}[ptb]
%\includegraphics[bb=119 272 488 567, width=9 cm]{g2t3.eps}\caption{(Color
%online) First order Approximate result compared with numerical result.
%Except $N$ other parameters are the same with Fig.~\ref{fig5}.}%
%\label{fig7}%
%\end{figure}
%\end{center}
%%%%%%%%%%%%%%%%%%%%%%%%%%%%%%%%%%%%%%%%%%%%%%%%%%%%%

\subsection{Reduced dynamics on the subspace with $m=0,1$ \label{subsection1}
}

When the photon number $n_{a}$ is in the vicinity of $1/2$, the
nearly degenerate energy levels $m=0,1$ ($\left\vert
j,0\right\rangle $ and $\left\vert j,1\right\rangle $) will be
strongly coupled with each other as a result of the driving, but
weakly coupled with other energy levels. Then the two
energy levels ($m=0,1$) form a relatively stable subsystem. Hence we
can treat the transitions from the subsystem to other levels by a
perturbative approach. In terms of the states with definite quantum
number $m$, the Hamiltonian $H^{(R)}=H_{0}+H_{I}$ can be decomposed in two
parts, the nonperturbative Hamiltonian,
\begin{eqnarray}
H_{0}&=&\omega_{0}\vert j,0\rangle \langle j,0\vert +\omega_{1}\vert
j,1\rangle \langle j,1\vert  \notag \\
&&+\Omega \sqrt{j( j+1) }\vert j,1\rangle
\langle j,0\vert +\text{h.c.}, \label{H0}
\end{eqnarray}
and the perturbation,
\begin{eqnarray}
H_{I} & =&\Omega_{2}\vert j,2\rangle \langle j,1\vert
+\Omega_{0}\vert j,0\rangle \langle j,-1\vert \notag
\\
& &+\sum_{m=-j,m\neq0,1}^{j}\omega_{m}\left\vert j,m\right\rangle
\langle j,m\vert  \notag \\
& &+\sum_{m=-j,m\neq-1,0,1}^{j-1}\Omega_{m+1}\vert j,m+1\rangle
\langle j,m\vert +\text{h.c.}.  \label{HI}
\end{eqnarray}

To see clearly if the atomic blockade effect occurs, namely, if it
is difficult to excite two more atomic excitations, we need to find the
transition amplitude for the system
initially prepared in the subspace spanned by $\vert j,0\rangle $
and $\vert j,1\rangle $ to the doubly excited state $|j,2\rangle$ around the critical point $n_{a}=1/2$. To
make $\vert j,0\rangle $ and $\vert j,1\rangle $ nearly degenerate,
we restrict $ 0<n_{a}<1$. We note that we can also choose any
other pair of nearly degenerate states around the corresponding
critical point which makes the pair nearly degenerate. We first
diagonalize the nonperturbative Hamiltonian~(\ref{H0}) as,
\begin{equation}
H_{0}=\lambda_{0}\left\vert \lambda_{0}\right\rangle \left\langle \lambda
_{0}\right\vert +\lambda_{1}\left\vert \lambda_{1}\right\rangle \left\langle
\lambda_{1}\right\vert.
\end{equation}
The two eigenstates are
\begin{equation}
\vert \lambda_{r}\rangle =A_{r}^{-1}\left[\xi_{r}\left\vert
j,0\right\rangle +\left\vert j,1\right\rangle \right],\hspace{0.5 cm} r=0,1,
\end{equation}
%%%%%%%%%%%%%%%%%%%%%%%%%%%%%%%%%%%%%%%%%%%%%%%%%%%%%%%
\begin{center}
\begin{figure}[ptb]
\includegraphics[bb=39 343 594 651, width=9.5 cm]{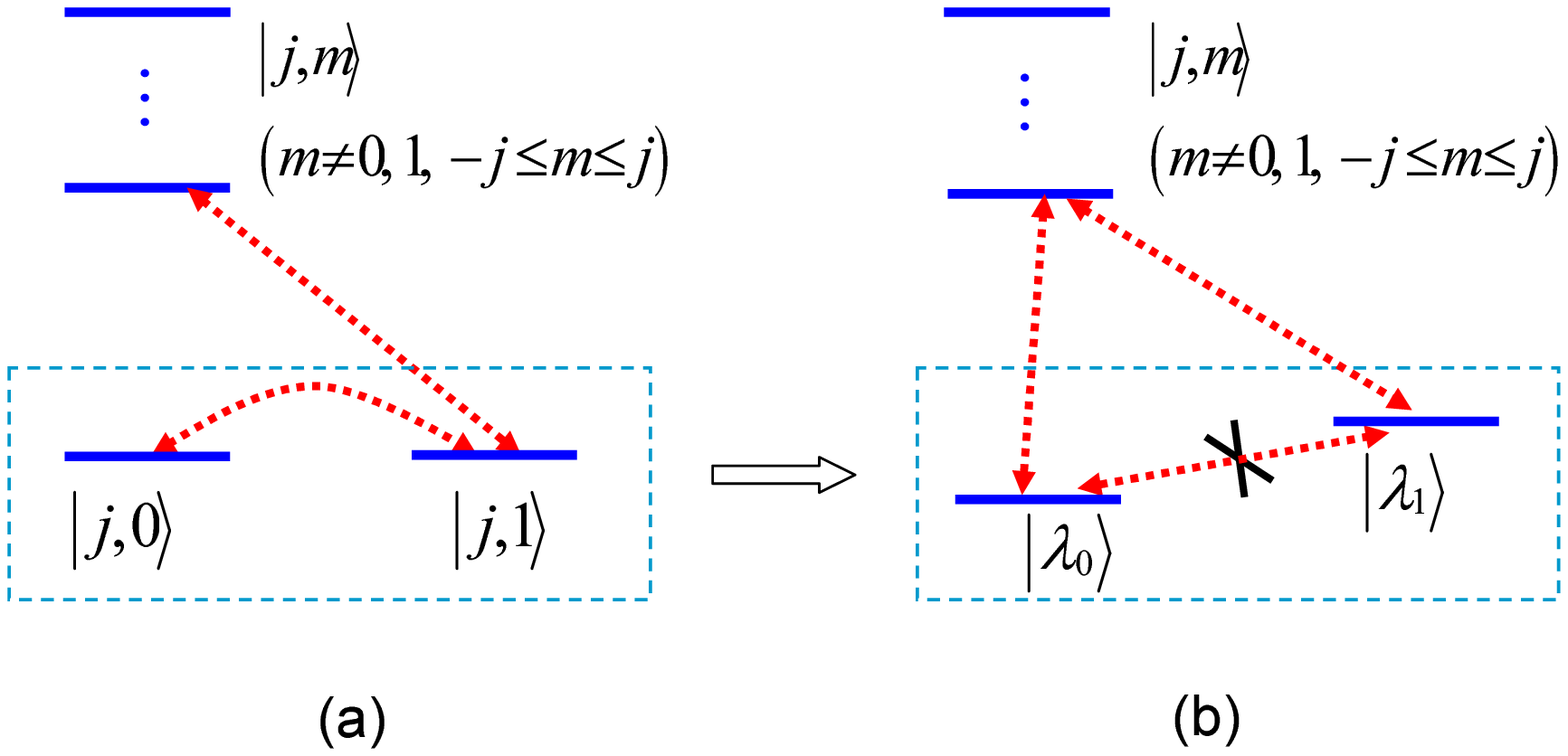}
\caption{(Color online) Energy-level diagram of the $m=0,1$ subsystem of the driven atomic ensemble. ($a$) The two nearly degenerate energy levels $|j,0\rangle$ and $|j,1\rangle$ are
strongly coupled with each other, when the average photon number in the cavity is
$n_{a}\approx1/2$, but weakly coupled with other energy levels.
($b$) The effective subsystem spanned by $|\lambda_{0}\rangle$ and
$|\lambda_{1}\rangle$ when using the perturbation approach.} \label{fig3}
\end{figure}
\end{center}
%%%%%%%%%%%%%%%%%%%%%%%%%%%%%%%%%%%%%%%%%%%%%%%%%%%%%%%
with corresponding eigenenergies,
\begin{eqnarray}
\lambda_{r} & =&jW+j^{2}W+W\delta+( -1)^{r+1}p \text{,}
\end{eqnarray}
where $A_{r}=\sqrt{\left\vert \xi_{r}\right\vert ^{2}+1}$ are
normalization constants with
\begin{equation}
\xi_{r}=-\frac{W\delta+(-1)^{r}p}{\Omega\sqrt{j\left( j+1\right) }},
\end{equation}
and
\begin{equation}
p\equiv\sqrt{W^{2}\delta^{2}+j\Omega^{2}+j^{2}\Omega^{2}}.
\end{equation}
We note that $\langle j,m\vert \lambda_{r}\rangle =0$ for
$m\neq0,1$. Therefore, $\left\vert \lambda_{0}\right\rangle $,
$\left\vert \lambda_{1}\right\rangle $ and $ \left\vert
j,m\right\rangle $ ($m\neq0,1$) form a complete basis of the
Hilbert space for a given $j$. In this basis, $H_{I}$ can be
expressed as,
\begin{eqnarray}
H_{I} & =&\Omega_{2}\left[\eta_{1}\vert j,2\rangle\langle
\lambda_{0}\vert +\eta_{2}\vert j,2\rangle \langle
\lambda_{1}\vert \right]  \notag \\
& &+\Omega_{0}\left[ \eta_{3}\left\vert \lambda_{0}\right\rangle
\left\langle j,-1\right\vert +\eta_{4}\left\vert
\lambda_{1}\right\rangle
\left\langle j,-1\right\vert \right]  \notag \\
& &+\sum_{m=-j,m\neq0,1}^{j}\omega_{m}\left\vert j,m\right\rangle
\left\langle j,m\right\vert  \notag \\
& &+\sum_{m=-j,m\neq-1,0,1}^{j-1}\Omega_{m+1}\left\vert
j,m+1\right\rangle \left\langle j,m\right\vert +\text{h.c.},
\label{Hi}
\end{eqnarray}
where
\begin{eqnarray}
\eta_{1}
&=&\frac{\xi_{1}A_{0}}{\xi_{1}-\xi_{0}},\hspace{0.5cm}\eta_{2}=
-\frac{\xi_{0}A_{1}}{\xi_{1}-\xi_{0}},  \notag \\
\eta_{3}
&=&\frac{A_{0}}{\xi_{0}-\xi_{1}},\hspace{0.5cm}\eta_{4}=-\frac{A_{1}
}{\xi_{0}-\xi_{1}},
\end{eqnarray}
which satisfy $\vert \eta _{1}\vert ^{2}+\vert \eta_{2}\vert ^{2}=1$
and $ \vert \eta_{3}\vert ^{2}+\vert \eta_{4}\vert ^{2}=1$. It
follows from Eq. (\ref{Hi}) that the transition between
$\left\vert \lambda_{0} \right\rangle $ and $\left\vert
\lambda_{1}\right\rangle $ is inhibited, which is shown in Fig.~\ref{fig3}. In order to calculate the
correlation function $g^{(2)}$ with the perturbed Hamiltonian, we move to the interaction picture
by choosing,
\begin{equation}
H_{0}^{\prime}=\lambda_{0}\left\vert \lambda_{0}\right\rangle
\left\langle \lambda_{0}\right\vert +\lambda_{1}\left\vert
\lambda_{1}\right\rangle \left\langle \lambda_{1}\right\vert
+\sum_{m=-j,m\neq0,1}^{j}\omega _{m}\left\vert j,m\right\rangle
\left\langle j,m\right\vert
\end{equation}
as the free Hamiltonian. In the interaction picture, the
Hamiltonian $ H^{(R)}=H_{0}^{\prime}+H_{I}^{\prime}$, where
\begin{equation}
H_{I}^{\prime}=\sum_{m=-j,m\neq0}^{j}\Omega_{m+1}\left\vert
j,m+1\right\rangle \left\langle j,m\right\vert +\text{h.c.}
\end{equation}
becomes
\begin{eqnarray}
V_{I}( t) & =&\Omega_{2}\left\vert j,2\right\rangle \left(
\eta_{1}\left\langle \lambda_{0}\right\vert e^{i\Delta_{2,0}t}+\eta
_{2}\left\langle \lambda_{1}\right\vert e^{i\Delta_{2,1}t}\right)  \notag \\
& &+\Omega_{0}\left( \eta_{3}\left\vert \lambda_{0}\right\rangle
e^{-i\Delta_{-1,0}t}+\eta_{4}\left\vert \lambda_{1}\right\rangle
e^{-i\Delta_{-1,1}t}\right) \left\langle -1,j\right\vert  \notag \\
& &+\sum_{m=-j,m\neq-1,0,1}^{j-1}\Omega_{m+1}\left\vert j,m+1\right\rangle
\left\langle m,j\right\vert e^{i\omega_{m+1,m}t}+\text{h.c.},  \notag \\
\end{eqnarray}
which is time-dependent. Here, we have defined
\begin{eqnarray}
\Delta_{m^{\prime},r}
&\equiv\omega_{m^{\prime}}-\lambda_{r}\text{,}\hspace{
0.3cm}\omega_{m,l}\equiv\omega_{m}-\omega_{l}\text{,}\hspace{0.3cm}
&
\end{eqnarray}
where $m^{\prime}\neq0,1\text{, }r=0,1$, and $\Delta_{m^{\prime},r}$ is the
energy difference between the diagonalized almost-degenerate energy
levels labeled by $\left\vert \lambda _{r}\right\rangle $ ($r=0,1$) and the other energy levels labeled by $\left\vert j,m\right\rangle $ ($m\neq0,1$).
%%%%%%%%%%%%%%%%%%%%%%%%%%%%%%%%%%%%%%%%%%%%%%%%%%%%%%%
%\begin{center}
%\begin{figure}[ptb]
%\includegraphics[bb=151 302 464 555, width=8 cm]{g2_delta1_2_1.eps}\caption{(Color
%online) Numerical result. $t=0$, $\tau=3$, $g_{0}=1000$, $g_{0}%
%/\Delta=-0.1$, $\delta_{c}=9$, other parameters are the same with
%fig.~\ref{fig5}.}%
%\label{fig10}%
%\end{figure}
%\end{center}
%%%%%%%%%%%%%%%%%%%%%%%%%%%%%%%%%%%%%%%%%%%%%%%%%%%%%%

\subsection{Reduced dynamics on the subspace with $m=j-1,j$}

Here we consider the effect of the QPT on the higher-order quantum
coherence around the critical point $n_{a}=j-1/2$. Similar to the
previous section, it can be seen that the states $
|j,j-1\rangle$ and $|j,j\rangle$ form a relative stable subsystem.
We can also treat the transitions from the subsystem ($m=j-1,j$) to other
energy levels by using a perturbative method. To this end we diagonalize the Hamiltonian in the subspace spanned by the two
nearly degenerate energy levels $\left\vert j,j-1\right\rangle $ and
$\left\vert j,j\right\rangle $. It follows from Eq.~(\ref{H}) that, the
nonperturbative Hamiltonian is,
\begin{eqnarray}
H_{0}^{c} & =&\omega_{j}\left\vert j,j\right\rangle \left\langle
j,j\right\vert +\omega_{j-1}\left\vert j,j-1\right\rangle \left\langle
j,j-1\right\vert  \notag \\
& &+\Omega_{j}\left\vert j,j\right\rangle \left\langle
j,j-1\right\vert +\text{h.c.}
\notag \\
& \equiv &\lambda_{0}^{c}\left\vert \lambda_{0}^{c}\right\rangle
\left\langle \lambda_{0}^{c}\right\vert +\lambda_{1}^{c}\left\vert
\lambda_{1}^{c}\right\rangle \left\langle \lambda_{1}^{c}\right\vert,
\end{eqnarray}
%%%%%%%%%%%%%%%%%%%%%%%%%%%%%%%%%%%%%%%%%%%%%%%%%%%%%%%%
\begin{figure}[ptb]
\includegraphics[bb=39 352 594 645, width=9 cm]{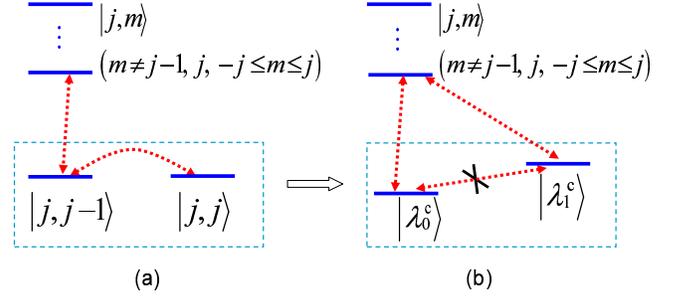}
\caption{(Color online) Energy-level diagram of the subsystem composed of
$m=j-1,j$ of the driven atomic ensemble. ($a$) The two
nearly degenerate energy levels $|j,j-1\rangle$ and $|j,j\rangle$
are strongly coupled with each other, when the average photon number in the cavity is
$n_{a}\approx j-1/2$, but weakly coupled with other energy levels.
($b$) The effective subsystem spanned by $|\lambda^{c}_{0}\rangle$
and $|\lambda^{c}_{1}\rangle$ by the perturbative approach.}
\label{fig4}
\end{figure}
%%%%%%%%%%%%%%%%%%%%%%%%%%%%%%%%%%%%%%%%%%%%%%%%%%%%%%
with the eigenenergies,
\begin{eqnarray}
\lambda_{r}^{c}  & =&2^{-1}\left[  -1-2n_{a}+4j(1+n_{a})\right]  W\nonumber\\
&&+2^{-1}
(-1)^{r+1}p^{c}, \hspace{0.8 cm}r=0,1,
\end{eqnarray}
and eigenvectors,
\begin{equation}
\left\vert \lambda_{r}^{c}\right\rangle =(A_{r}^{c})^{-1}\left[\xi_{r}^{c}\left\vert
j,j-1\right\rangle +\left\vert j,j\right\rangle\right],\hspace{0.5 cm} r=0,1,
\end{equation}
where $A_{r}^{c}=\sqrt{\left\vert \xi_{r}^{c}\right\vert ^{2}+1}$ ($r=0,1$) are
normalization constants with,
\begin{eqnarray}
\xi_{r}^{c}=\left[  2\Omega\sqrt{2j}\right]  ^{-1}\left[  (-1+2j-2n_{a}%
)W+(-1)^{r+1}p^{c}\right],
\end{eqnarray}
and
\begin{equation}
p^{c}\equiv\sqrt{(1-2j+2n_{a})^{2}W^{2}+8j\Omega^{2}}.
\end{equation}
Similar to the above Subsec.~\ref{subsection1}, we also note that $\langle j,m\vert \lambda^{c}_{r}\rangle =0$ for
$m\neq j-1,j$. Therefore, $\left\vert \lambda^{c}_{0}\right\rangle $,
$\left\vert \lambda^{c}_{1}\right\rangle $ and $ \left\vert
j,m\right\rangle $ ($m\neq j-1,j$) form a compete basis of the
Hilbert space for a given $j$.
 In terms of $\left\vert
\lambda_{0}^{c}\right\rangle $ and $\left\vert \lambda_{1}^{c}\right\rangle $%
, the residual terms of the Hamiltonian~(\ref{H})
$H_{I}^{c}=H^{(R)}-H_{0}^{c}$ read as
\begin{eqnarray}
H_{I}^{c}&=&\Omega_{j-1}\left[ \left( \eta_{3}^{c}\left\vert
\lambda_{0}^{c}\right\rangle +\eta_{4}^{c}\left\vert
\lambda_{1}^{c}\right\rangle \right) \left\langle j,j-2\right\vert
+\text{h.c.}
\right]  \notag \\
& &+\sum_{m=-j}^{j-2}\omega_{m}\left\vert j,m\right\rangle \left\langle
j,m\right\vert  \notag \\
&&+\sum_{m=-j}^{j-3}\Omega_{m+1}\left( \left\vert j,m+1\right\rangle
\left\langle j,m\right\vert +\text{h.c.}\right) \label{Hi_c},
\end{eqnarray}
where we used the expressions,
\begin{equation}
\left\vert j,j\right\rangle =\eta_{1}^{c}\left\vert
\lambda_{0}^{c}\right\rangle +\eta_{2}^{c}\left\vert
\lambda_{1}^{c}\right\rangle \text{, }\left\vert j,j-1\right\rangle
=\eta_{3}^{c}\left\vert \lambda_{0}^{c}\right\rangle +\eta_{4}^{c}\left\vert
\lambda_{1}^{c}\right\rangle \text{,}
\end{equation}
with coefficients defined by
\begin{eqnarray}
\eta_{1}^{c} & =&\frac{\xi_{1}^{c}A_{0}^{c}}{\xi_{1}^{c}-\xi_{0}^{c}}
,\hspace{0.5 cm}\eta_{2}^{c}=-\frac{\xi_{0}^{c}A_{1}^{c}}{\xi_{1}^{c}-\xi_{0}},  \notag \\
\eta_{3}^{c} & =&\frac{A_{0}^{c}}{\xi_{0}^{c}-\xi_{1}^{c}},\hspace{0.5 cm}\eta_{4}^{c}=
-\frac{A_{1}^{c}}{\xi_{0}^{c}-\xi_{1}^{c}},
\end{eqnarray}
which satisfy $\left\vert \eta_{1}^{c}\right\vert ^{2}+\left\vert
\eta_{2}^{c}\right\vert ^{2}=1$, $\left\vert \eta_{3}^{c}\right\vert
^{2}+\left\vert \eta_{4}^{c}\right\vert ^{2}=1$. It
follows from Eq.~(\ref{Hi_c}) that, there is no transition between
$\left\vert \lambda^{c}_{0} \right\rangle $ and $\left\vert
\lambda^{c}_{1}\right\rangle $, which is shown in Fig.~\ref{fig4}.
In order to change to the interaction picture, we choose the diagonalized terms,
\begin{equation}
H_{0}^{c\prime}=\lambda_{0}^{c}\left\vert
\lambda_{0}^{c}\right\rangle \left\langle \lambda_{0}^{c}\right\vert
+\lambda_{1}^{c}\left\vert \lambda _{1}^{c}\right\rangle
\left\langle \lambda_{1}^{c}\right\vert +\sum
_{m=-j}^{j-2}\omega_{m}\left\vert j,m\right\rangle \left\langle
j,m\right\vert
\end{equation}
as the free Hamiltonian, and the corresponding interaction Hamiltonian,
\begin{equation}
H_{I}^{c\prime}=\sum_{m=-j}^{j-2}\Omega_{m+1}\left( \left\vert
j,m+1\right\rangle \left\langle j,m\right\vert +\text{h.c.}\right).
\end{equation}

Finally, we obtain the interaction Hamiltonian,
\begin{eqnarray}
V_{I}^{c}\left( t\right) & =&\Omega_{j-1}\left( \eta_{3}^{c}\left\vert
\lambda_{0}^{c}\right\rangle
e^{-i\Delta_{j-2,0}^{c}t}+\eta_{4}^{c}\left\vert
\lambda_{1}^{c}\right\rangle e^{-i\Delta_{j-2,1}^{c}t}\right) \left\langle
j,j-2\right\vert  \notag \\
& &+\sum_{m=-j}^{j-3}\Omega_{m+1}\left\vert j,m+1\right\rangle
\left\langle j,m\right\vert e^{i\omega_{m+1,m}t}+h.c
\end{eqnarray}
in the interaction picture where
\begin{align}
\Delta_{m^{\prime},r}^{c} & \equiv\omega_{m^{\prime}}-\lambda_{r}^{c}\text{,
}\hspace{0.3cm}\omega_{m,l}\equiv\omega_{m}-\omega_{l}\text{, }\hspace{0.3cm}
&
\end{align}
for $m^{\prime}\neq j-1,j\text{, }r=0,1$ and $\Delta^{c}_{m^{\prime},r}$ is the
energy difference between the diagonalized almost degenerate energy
levels labeled by $\left\vert \lambda^{c} _{r}\right\rangle $ ($r=0,1$) and the other energy levels labeled by $\left\vert j,m\right\rangle $ ($m\neq j-1,j$).

Note that Figs.~\ref{fig3} and~\ref{fig4} show transitions between three level systems, where some of the transitions are turned on and off. Indeed, it is also possible to turn on and off transitions between three energy levels using artificial atoms made of superconducting qubits~\cite{3-transitions}.

\section{Statistical properties of the atomic excitations}

Since $J_{-}$ ($J_{+}$) can decrease (increase) a single excitation, their roles are similar to the actions of the annihilation
(creation) operator of photons $a$ ($a^{\dag}$) for the usual
bosonic system. Using the Holstein-Primakoff
transformation~\cite{HP},  the angular momentum
operators can be expressed in terms of a single bosonic mode,
\begin{align}
J_{+} =b^{\dag}\sqrt{N-b^{\dag}b},\hspace{0.2CM} J_{-} =\left(\sqrt{N-b^{\dag}b}\right)b,%
\hspace{0.2CM} J_{z} =b^{\dag}b-\frac{N}{2}.
\end{align}
The angular momentum operators will become bosonic operators in the
limit of large $N$ and low excitations, namely, $\langle b^{\dag}b\rangle\ll N$~\cite{JFHuang}. Specifically, in this condition, one can expand the square term $\sqrt{N-b^{\dag}b}$ on the order of $(b^{\dag}b)/N$ and keep to the zeroth order of $b^{\dag}b/N$. Then it is straightforward to see that $J_{+}\simeq b^{\dag}\sqrt{N}$ and $J_{-}\simeq \sqrt{N}b$~\cite{JFHuang}. Then we can define a generalized second-order coherence function,
\begin{equation}
g^{\left( 2\right) }\left( \tau,t\right) =\frac{\left\langle
J_{+}\left( t\right) J_{+}\left( t+\tau\right) J_{-}\left(
t+\tau\right) J_{-}\left( t\right) \right\rangle }{\left\langle
J_{+}\left( t\right) J_{-}\left( t\right) \right\rangle \langle
J_{+}( t+\tau) J_{-}( t+\tau)\rangle }, \label{g_2_t}
\end{equation}
for the symmetric collective excitations of the atomic ensemble,
which can be regarded as a normalized correlation function. Please note that this definition is in normal order on the angular momentum operators $J_{+}$ and $J_{-}$, which satisfy that the average of any analytical function of $J_{+}J_{-}$ in normal order over the ground state $|j,-j\rangle$ equals zero, that is, $\langle j,-j\vert :f(J_{+}J_{-}):\vert j,-j\rangle =0$. Here $j$ is a conserved quantity. This property satisfies the
conventional normal order definition about the bosonic operators  $\langle 0\vert :f(b^{\dag}b):\vert0\rangle =0$ in the second coherence function. This
coherence function $g^{(2)}(\tau,t)$ is proportional to the joint probability of observing one excited atom at time $
t$ and another one at time $t+\tau$. To study the generalized second-order coherence function $g^{(2)}(\tau,t)$ in the stationary state, below we consider it in a unitary evolution case (without dissipation) and also in a dissipation case but at a steady state.

%However, when $N$ is finite, the generalized second-order coherence function can't reduce to conventional bosonic seconder order coherence function. In this case, the generalized second-order coherence function can't used to discuss the statistical properties of atomic collective excitations such as super-Possion and sub-Possion statistics as conventional coherence function.

\subsection{The case without dissipation}

Firstly, we consider the case where the system is free of dissipation. In
this case, $ \langle\cdots\rangle$ in Eq.~(\ref{g_2_t}) for $g^{(2)}(\tau,t)$ denotes the
average of an observable over the initial pure state,
\begin{equation}
\left\vert \psi\left( 0\right) \right\rangle =\sum_{m=-j}^{j}c_{m}\left\vert
j,m\right\rangle,   \label{psi0}
\end{equation}
where $\sum^{j}_{m=-j}|c_{m}|^{2}=1$.

\begin{center}
\begin{figure}[ptb]
\includegraphics[bb=125 290 436 545, width=8 cm]{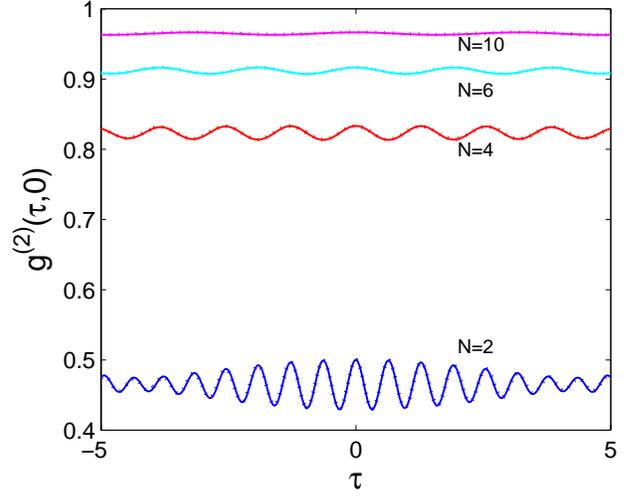}
\caption{(Color online) Second-order correlation function $g^{(2)}(\tau,0)$ versus the time interval
$\tau$ for $N=2$ (blue curve), $N=4$ (red curve), $N=6$ (cyan curve), and
$N=10$ (magenta curve), respectively, in the case without dissipation. Recall that $g^{(2)}(\tau,0)$ is proportional to the joint probability of observing one excited atom at time $t=0$ and another one at time $\tau$.
The first-order approximate results are shown using dashed curves and the exact
numerical results are shown using solid curves. The dashed curves overlap with the solid curves very well. Other parameters are
$g_{0}=100$, $ g_{0}/\Delta=-0.1$, $\Omega N=0.1|W|$,
$\delta=-0.02$.} \label{fig5}
\end{figure}
\end{center}

%%%%%%%%%%%%%%%%%%%%%%%%%%%%%%%%%%%%%%%%%%%%%%%%%%%%%%%
%\begin{center}
%\begin{figure}[ptb]
%\includegraphics[bb=51 226 533 609, width=6cm]{g_2_0.eps}\caption{(Color
%online) Numerical result with dissipation in steady state. $N=2$ (blue solid line),
%$N=3$ (red dashed line), $N=5$ (cyan dash-dot line), $N=10$ (magenta dotted line).
%$\gamma=1$, $g_{0}=100$, $g_{0}/\Delta=-0.1$, $\Omega N=0.1W$.}%
%\label{fig6}%
%\end{figure}
%\end{center}
%%%%%%%%%%%%%%%%%%%%%%%%%%%%%%%%%%%%%%%%%%%%%%%%%%%%%%%
We next calculate the generalized second-order coherence function around the point
$\delta=0$ (i.e., $n_{a}=1/2$). Since $U(\tau)=U_{0}(\tau)U_{I}(\tau)
$, where $U_{0}(\tau) =\exp(-iH_{0}^{\prime}\tau)$ and
$U_{I}(\tau)=T\exp{ [-i\int_{0}^{\tau}V_{I}(\tau
^{\prime})d\tau^{\prime}]}$ are the free evolution and
the dynamics due to the interaction, respectively. We note that
$U_{0}^{\dag}(\tau) J_{+}J_{-}U_{0}(\tau) =J_{+}J_{-}$ is useful in
the following calculations. Using this result, the generalized second-order coherence function
$g^{\left( 2\right) }\left( \tau,0\right) $ becomes
\begin{equation}
g^{(2)}(\tau,0) =\frac{\langle \psi^{\prime }(0)\vert
U_{I}^{\dag}(\tau)J_{+} J_{-}U_{I}(\tau) \vert \psi^{\prime}(0) \rangle}{%
\langle \psi^{\prime}(0)\vert \psi^{\prime}(0)\rangle \langle \psi(0) \vert
U_{I}^{\dag}(\tau)J_{+}J_{-}U_{I}(\tau) \vert \psi(0) \rangle },
\label{g_2_0}
\end{equation}
where $\vert \psi^{\prime}(0)\rangle =J_{-}|\psi(0)\rangle.$ We will
calculate analytically the generalized second-order coherence function by applying standard
perturbation theory, with $V_{I}(t)$ as a perturbation. Let us first
consider the conditions where the perturbation approach is valid.
If we tune the atom-field detuning $\Delta$ and the Rabi frequency $\Omega$ of the
driving field to be suitable and place an appropriate number of
atoms in the cavity, we can make the perturbation theory valid,
that is,
%\begin{equation}
%\Omega N\ll\Delta_{m^{\prime},j}\text{, \ \ }\Omega N\ll\omega_{m+1,m}
%\end{equation}
for $m^{\prime}=2,-1$, $r=0,1,$ and $m\neq-1,0,1$ these conditions
explicitly are
\begin{eqnarray}
\Omega_{2}\eta_{1} & \ll&\Delta_{2,0},\hspace{0.6 cm}\Omega_{2}\eta_{2}\ll
\Delta_{2,1},  \notag \\
\Omega_{0}\eta_{3} & \ll&\Delta_{-1,0},\hspace{0.5 cm}\Omega_{0}\eta_{4}\ll
\Delta_{-1,1},  \notag \\
\Omega_{m+1} & \ll&\omega_{m+1,m}.  \label{condition 1}
\end{eqnarray}
Under the above conditions, we can treat the time-evolution operator $%
U_{I}\left( \tau\right) $ perturbatively. When $n_{a}$ is in the
vicinity of the critical point $n_{a}^{c}$ (for $m=0,1$, $n_{a}^{c}=1/2$%
), the energy levels of $\left\vert \lambda_{0}\right\rangle $ and $%
\left\vert \lambda _{1}\right\rangle $ are nearly degenerate. The
energy differences $\Delta_{i,j}$ and $\omega_{m+1,m}$ ($m\neq-1, 0,
1$) are very large compared with the level spacing between
$\left\vert \lambda_{0}\right\rangle $ and $\left\vert
\lambda_{1}\right\rangle $. Hence, under this constraint, the above
conditions~(\ref{condition 1}) can be satisfied by varying the Rabi
frequency $\Omega$. Since the state $\vert j,0\rangle $ is the
ground state when $0<n_{a}<1/2$, then $\vert j,1\rangle $ is the state by
exciting one more atom. Similarly, $\vert j,2\rangle $ has two more
excitations than the ground state, and has a much higher
energy than that of $\vert j,1\rangle $. However, $\vert
j,1\rangle $ is the ground state when $1/2<n_{a}<1$, yet $\vert
j,0\rangle $ is an excited state which has one less atomic excitation
than the ground state $\vert j,1\rangle $. To see if two
excitations are suppressed, we choose $c_{0}=c_{1}=1/\sqrt{2}$ and
$c_{m}=0$ (for $m\neq0,1$) in the initial state,
\begin{equation}
\vert \psi( 0) \rangle =c_{0}\vert j,0\rangle +c_{1}\vert j,1\rangle=\frac{1}{\sqrt{2}}(\vert j,0\rangle +\vert j,1\rangle).
\end{equation}
When the average photon number is in the vicinity of $n_{a}^{c}=1/2$, the states $\vert j,0\rangle $ and $\vert j,1\rangle $ are
nearly degenerate. Notice that here the average photon number is in the domain of
$0<n_{a}<1$ and around $ n_{a}^{c}=1/2$, that is, $-1/2<\delta<1/2$, and
$\vert \delta\vert $ is very small. Using first-order perturbation theory, the generalized second-order coherence function in
Eq.~(\ref{g_2_0}) is approximately
\begin{equation}
g^{(2)}(\tau,0)\simeq\frac{X}{( j+1)jY},  \label{g2}
\end{equation}
where
\begin{eqnarray}
X & \equiv& x_{1}+x_{2}+x_{3}+x_{4}+x_{5},  \notag \\
Y & \equiv& y_{1}+y_{2}+y_{3}.
\end{eqnarray}
The parameters $x_{\ell_{1}}$ ($\ell_{1}=1,2,\cdots,5$) and $y_{\ell_{2}}$
($\ell_{2}=1,2,3$) have complicated expressions, which are presented in the
appendix. The generalized second-order coherence function given by Eq.~(\ref{g2}) is illustrated in
Fig.~\ref{fig5}. It is shown that, as $N$ increases, the value of $g^{(2)}(\tau,0)$ approaches unity with some oscillations. Physically, Eq.~(\ref{g2}) describes the joint
probability of observing one excited atom at instant $t=0 $ and
another after a time interval $\tau$. In Sec.~\ref{V}, we use Eq.~(\ref{g2}) to analyze the atomic blockade effect.
%%%%%%%%%%%%%%%%%%%%%%%%%%%%%%%%%%%%%%%%%%%%%%%%%%%%%%
%%%%%%%%%%%%%%%%%%%%%%%%%%%%%%%%%%%%%%%%%%%%%%%%%%%%%%

\subsection{the case with dissipation}

In this subsection, we consider the system surrounded by a thermal
reservoir at zero temperature. When the system is prepared in a state
with density operator $\rho_{s}$, the generalized second-order coherence function is written explicitly as
\begin{equation}
g^{(2)}(\tau,t)=\frac{\text{Tr}[ J_{+}J_{+}(\tau)J_{-}(\tau)
J_{-}\rho_{s}(t)]}{\text{Tr}[ J_{+}J_{-}\rho_{s}(t)]\text{Tr}[
J_{+}J_{-}\rho_{s}(t+\tau)]}.  \label{g_2_tao_decay}
\end{equation}
According to Eq.~(\ref{g_2_tao_decay}), we need to calculate the
time-dependent density operator $\rho_{s}(t)$ of the atoms. In the
regime of weak coupling of the driving field~\cite{Breuer}, which
demands the driving field to only perturbatively change the energy
levels, and assuming the atomic ensemble to be in a common reservoir,
then the master equation is approximately
\begin{equation}
\frac{d\rho_{s}\left( t\right) }{dt}=-i\left[ H^{(R)},\rho_{s}\left(
t\right) \right] +\gamma\left[ J_{-}\rho_{s}\left( t\right)
J_{+}-\frac{1}{2}\left\{ J_{+}J_{-},\rho_{s}\left( t\right) \right\}
\right],
\end{equation}
where $\gamma$ is the collective decay rate of the atomic ensemble. Since the photon number is a conserved number, and the frequency of photons is in large detuning, it does not influence the dynamical evolution of the atoms. Then the influence of the decay of the photons is negligibly small to the atoms.
We resort to numerical calculations to show the results about the
steady state by plotting $g^{(2) }(0,t\rightarrow\infty)$ versus
$\delta$ in Fig.~\ref{fig6}~(a) and
$g^{(2)}(\tau,t\rightarrow\infty) $ versus $\tau$ in
Figs.~\ref{fig6}~(b)-(d). By comparing them with the results in
Fig.~\ref{fig5}, we will discuss them in the next section.

\section{Double excitation Effects I: The Atomic Blockade Effect\label{V}}

In this and the next section, we discuss some physical effects
due to the double collective excitation, according to their quantum
statistics characterized by the generalized second-order coherence function $g^{(2)}(\tau,t)$ introduced in the last section.
We have calculated the generalized second-order coherence function in the above section both in the
dissipation-free case and also the case with dissipation. In this
section, we discuss the results in both cases according to
the above calculations. We illustrate the analytical results~(\ref{g2}) and compare them with
the numerical results by plotting in Fig.~\ref{fig5} the generalized second-order coherence function $g^{(2) }(\tau,0)
$ versus the time interval $\tau$ around $\delta_{c}=0$, without
dissipation. The generalized second-order coherence function is plotted for $N=2, 4, 6,
10$ atoms, respectively. It is clear from Fig.~\ref{fig5} that, close to the
critical point $\delta_{c}=0$, our analytical approximate
results~(\ref{g2}) (dashed line) agree very well with the numerical
result~(\ref{g_2_t}) (solid line). Obviously, $g^{(2)}(\tau,0)<1$ at any time
interval $\tau$. This shows that the atomic collective symmetric
excitations obey sub-Poissonian statistics. It can also be found that
as $N$ increases, $ g^{(2)}(\tau,0)<1$ oscillates slower and slower
and approaches unity, especially for
$g^{(2)}(0,0)$. That is because the generalized second-order coherence function at $ \tau=0$ is
\begin{equation}
g^{\left( 2\right) }\left( 0,0\right) =1-\frac{4}{N^{2}+2N}.
\label{g_2_0_0}
\end{equation}
Hence $g^{\left( 2\right) }\left( 0,0\right) $ increases as $N$ increases.
In the thermodynamic limit $N\rightarrow\infty$,
\begin{equation}
g^{(2)}(0,0) \rightarrow1.
\end{equation}
This shows that when $N$ is smaller, the effect of suppressing the doubly-excited state is enhanced.

%%%%%%%%%%%%%%%%%%%%%%%%%%%%%%%%%%%%%%%%%%%%%%%%%%%%%%%
\begin{figure}[!htb]
\centering
\psfig{file=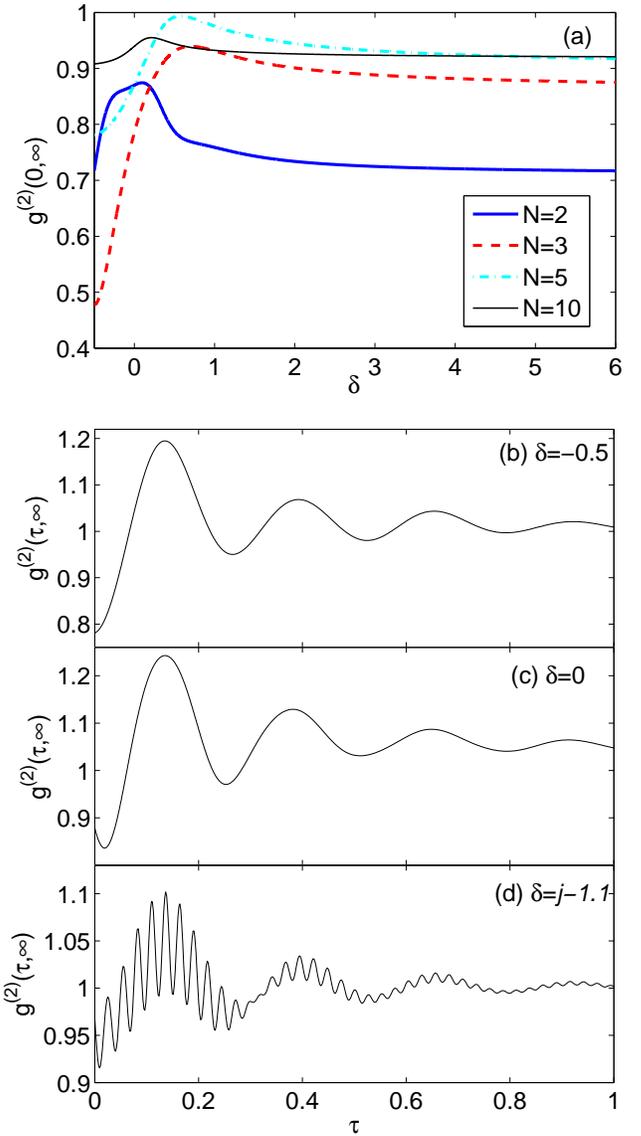,bb=101 46 509 791,width=8.2 cm,origin=br,angle=0}  %\hspace{0.5in}
\caption{(Color online) Numerical results for the generalized second-order coherence function $g^{(2)}(0,\infty)$ and $g^{(2)}(\tau,\infty)$ with dissipation in
the steady state. (a) $g^{(2)}(0,\infty)$ versus $\protect\delta$
for $N=2$ (blue thick solid curve), $N=3$ (red dashed curve), $N=5$ (cyan
dashed-dotted curve) and $N=10$ (black thin solid curve), respectively; (b)-(d) $g^{(2)}(\tau,\infty)$ versus $%
\tau$ with $N=5$ for $\delta=-0.5$, $\delta=0$, and $%
\delta=j-1.1$, respectively. Other common parameters are $
\gamma=1$, $g_{0}=100$, $g_{0}/\Delta=-0.1$, and $\Omega N=0.1|W|$.}
\label{fig6}
\end{figure}
%%%%%%%%%%%%%%%%%%%%%%%%%%%%%%%%%%%%%%%%%%%%%%%%%%%%%%%
Figure~\ref{fig6} shows the results for the dissipative case.
Figure~\ref{fig6}(a) shows $g^{(2)}(0,\infty)$ versus the average photon number
$\delta$ in steady state for $N=2, 3, 5, 10$ atoms, respectively.
As shown in this figure, the value of $g^{(2)}(0,\infty)$ increases
as $N$ increases for a larger average photon number. For a definite $N$ and a small
value of $\delta$, $g^{(2)}(0,\infty)$ increases as $ \delta$
increases. At some intermediate time there is a peak in $g^{(2)}(0,\infty)$ followed
 by a steady decrease, asymptotically approaching a constant value for large $\delta$.
%Then in the right hand side of the peak, it deceases to a
%steady value for a larger $\delta$.
 The smallest value of $g^{(2)}(0,\infty)$ occurs at $\delta=-0.5$. This phenomenon is also
prominent in Figs.~\ref{fig6}(b)-(d).
Figures~\ref{fig6}(b)-(d) show $ g^{(2)}(\tau,\infty)$ versus
$\tau$ for $N=5$ and $\delta=-0.5$, $0$ and $j-1.1$, respectively.
The antibunching effect of collective excitations of an atomic
ensemble is observed since the envelop of $g^{(2)}( \tau,\infty)$ shows $g^{(2)}(0,\infty)<g^{(2)}( \tau,\infty)$ with some increasingly rapid oscillations as $\delta$ increases in Figs.~\ref{fig6}(b)-(d). Additionally we note that $
g^{(2)}(\tau,\infty)$ approaches one, as expected, after some
oscillations. This indicates the probability of two collective excitations of the atomic ensemble
at the same time ($\tau=0$) is smaller than that at a different time
($\tau\neq0$). Therefore, the resonant excitations from the ground
state to the doubly excited state are suppressed. This is a clear
signature of the atomic blockade. Compared with Fig.~\ref{fig5},
this result is better and closer to physical reality. As shown
in Fig.~\ref{fig5}, the generalized second-order coherence function only oscillates with time interval
$\tau$ and does not approach $1$ as we expect when $
\tau\rightarrow\infty$. In Ref.~\cite{Ficek1984}, the photon antibunching effect is also obtained in only two interacting atoms. However, the antibunching effect we obtain is about atomic collective excitations, and the photon number is a conserved number. In this sense, we do not need to consider the photon correlation.

To conclude this section, we give some remarks about the atomic
blockade. For applications in quantum information, the atomic
blockade provides a novel approach to physical implementation of
scalable quantum logic gates such as implementing a CNOT gate between two
atoms~\cite{Rydberg blockade1,Rydberg blockade2,Rydberg blockade3}
and some kinds of quantum protocols~\cite{Rydberg blockade
protocal1,Rydberg blockade protocal2,Rydberg blockade
protocal3,Rydberg blockade protocal4}. Furthermore, as double excitation are inhibited in the Rydberg blockade mechanism, it also
supplies a fascinating approach to store quantum
information~\cite{Rydberg blockade1,Rydberg blockade2}. However, the
dipole-dipole interaction depends on the distance between Rydberg
atoms. To achieve a stronger interaction, it requires the atoms to
be closer in space or to be excited to higher Rydberg states, in
which the principal quantum number is very large, but this will not
be convenient to control the atoms individually~\cite{Rydberg
blockade1,Rydberg blockade2,Rydberg blockade3}. Such as the Rydberg
levels $n=79$ and $90$, the corresponding blockade shift is
$2\pi\times3$ and $2\pi\times9.5$~MHz at an interatom distance
$10.2$~$\mu$m, respectively. To achieve a larger energy-level shift
due to the Rydberg blockade, the distance needs to be decreased, and
thus the coherent manipulation of individual atoms is difficult. It is this
consideration that motivates us to find a new mechanism inducing a
stronger interatom coupling, valid for long distances and
controllable to improve the dipole-dipole interaction. We note that
in Ref.~\cite{strong coupling}, the coupling strength between atom
and photons can reach $2\pi\times120$ MHz in a high-finesse cavity,
which leads us to anticipate that the strong atom-photon coupling will
induce a stronger interatom interaction among atoms. In addition, this
interaction can be feasibly controlled by the volume of high-finesse
microcavities. This fact means that to achieve a strong interatom interaction among atoms will not take stringent
requirements on manipulating atoms individually. Therefore, from the point of view of the controllability and strength of the interaction, the photon-induced interaction among atoms in our system is better than the dipole-dipole interaction inducing the Rydberg blockade.

\section{Double excitation Effects II: Sensitivity of The Quantum Phase Transition}

As the system %effectively described by the LMG model
 possesses QPTs,
we now analyze how to control the QPT by photons in the cavity. To
show the effect of the QPT on $g^{\left( 2\right) }\left(
\tau,0\right) $ more clearly, we consider
the $g^{\left( 2\right) }\left( \tau,0\right) $ around the critical point $%
n_{a}^{c}=j-1/2$ at a fixed time interval $\tau$. Then, according to Eq.~(%
\ref{psi0}) we choose $c_{j-1}=c_{j}=1/\sqrt{2}$ and $c_{m}=0$ ($m\neq j-1,j$%
) in the initial state, namely
\begin{equation}
\vert \psi^{c}(0)\rangle =\frac{1}{\sqrt{2} }\vert j,j-1\rangle +\frac{1}{%
\sqrt{2}}\vert j,j\rangle.  \label{initial 2}
\end{equation}
With the relations $U\left( \tau\right) =U_{0}^{c}\left( \tau\right)
U_{I}^{c}\left( \tau\right)$ for $U_{0}^{c}(\tau) =\exp(iH_{0}^{c\prime}\tau)$%
, $U_{I}^{c}(\tau)
=T\exp[-i\int_{0}^{\tau}V_{I}^{c}(\tau^{\prime})d\tau^{\prime}]$, it
follows from Eq.~(\ref{g_2_t}) that
\begin{eqnarray}
&&U_{0}^{c\dag}(\tau)J_{+}J_{-}U_{0}^{c}(\tau)\notag \\
&=&i\alpha(
t)(\vert j,j-1\rangle \langle j,j\vert -\vert j,j\rangle \langle j,
j-1\vert )  \notag \\
& &+\beta(t) ( \vert j,j-1\rangle \langle j,j\vert +\vert j,j\rangle
\langle j,j-1\vert )
\notag \\
& &+\gamma(t) ( \vert j,j\rangle\langle j,j\vert -\vert j,j-1\rangle
\langle j,j-1\vert ) \notag \\
&&+J_{+}J_{-}.
\end{eqnarray}
The explicit expressions of the coefficients $\alpha\left( t\right) $,
$\beta(t)$ and $\gamma\left( t\right) $\ are given in the appendix.

Next, we use the perturbation approach to calculate the generalized second-order coherence function under the
following conditions
%\begin{eqnarray}
%\Omega\sqrt{N} & \ll&\Omega N\ll\Delta_{m^{\prime},j}^{c}  \notag \\
%\Omega\sqrt{N} & \ll&\Omega N\ll\omega_{m+1,m}.
%\end{eqnarray}
for $m^{\prime}=j-2$, $r=0,1$, and $m\neq j-2,j-1,j$, %these
%conditions explicitly are
\begin{eqnarray}
\Omega_{j-1}\eta_{3}^{c} & \ll &
\Delta_{j-2,0}^{c}\text{,}\hspace{0.5cm}
\Omega_{j-1}\eta_{4}^{c}\ll\Delta_{j-2,1}^{c},  \notag \\
\Omega_{m+1} & \ll & \omega_{m+1,m}\text{.}
\end{eqnarray}
Up to first order in $V_{I}^{c}\left( \tau\right) $, we obtain
\begin{equation}
g^{\left( 2\right) }\left( \tau,0\right) \simeq\frac{
\sum^{7}_{\ell_{1}=1}x_{\ell_{1}}^{c}}{%
\left( 3j-1\right) \left( \sum^{4}_{\ell_{2}=1}y_{\ell_{2}}^{c}\right) },
   \label{g2QPT}
\end{equation}
where the parameters $x_{\ell_{1}}^{c}$ ($\ell_{1}=1,2,\cdots,7$) and $y_{\ell_{2}}^{c}$ ($%
\ell_{2}=1,2,3,4$) have very long expressions, so we give these in the
appendix.

\begin{figure}[ptb]
\includegraphics[bb=74 118 499 745, width=8.2 cm]{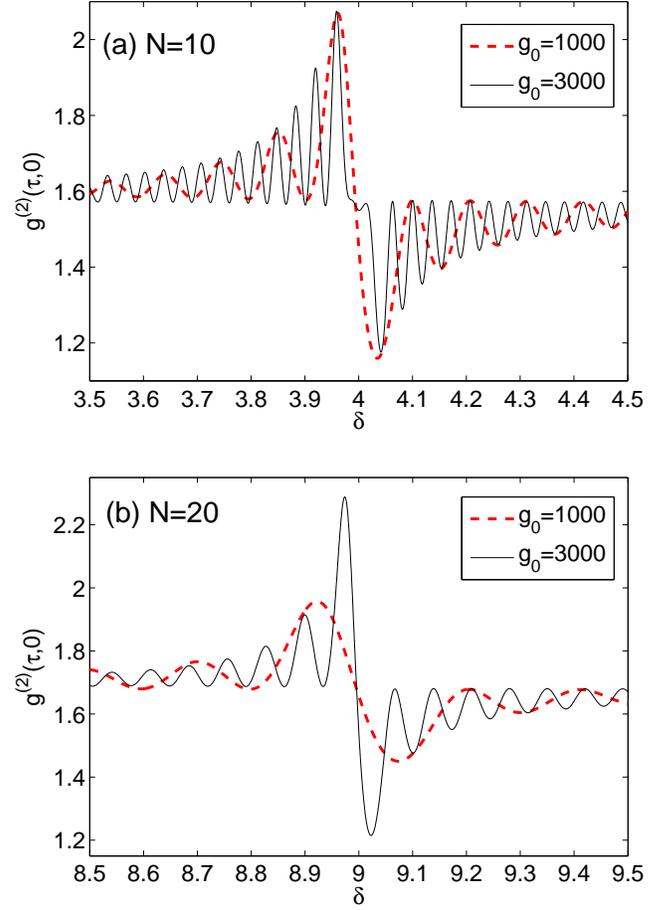}
\caption{(Color online) Numerical results for the generalized second-order coherence function $g^{(2)}(\tau,0)$. Here: $t=0$,
$\tau=3$, $ g_{0}=1000$ (red thick solid curve), $g_{0}=3000$ (black thin
solid curve), (a) $N=10$, $ \delta_{c}=4$, (b) $N=20$,
$\delta_{c}=9$; other parameters are the same as in
Fig.~\ref{fig5}.} \label{fig7}
\end{figure}

We also numerically calculate the generalized second-order coherence function in Eq.~(\ref{g_2_t}), and
then plot $g^{(2)}(\tau,0)$ versus $\delta$ in
Fig.~\ref{fig7}. As Fig.~\ref{fig7} indicates, the statistical coherence of atomic excitations is very sensitive to
%%%%%%%%%%%%%%%%%%%%%%%%%%%%%%%%%%%%%%%%%%%%%%%%%%%%%%
%%%%%%%%%%%%%%%%%%%%%%%%%%%%%%%%%%%%%%%%%%%%%%%%%%%%%%
the critical point $n^{c}_{a}=j-1/2$. The probability of double atomic excitation is above the dotted straight line in the left-hand side of the critical point, while it is below this curve in the right-hand side of the critical point. Furthermore the envelope exponentially
decays. When the average photon number is in the domain of~$j-1\leq n_{a}\leq j-1/2$
($j-3/2\leq\delta\leq j-1$), the energy level of $\left\vert
j,j\right\rangle $ is higher than $ \left\vert j,j-1\right\rangle $
but lower than $\left\vert j,j-2\right\rangle $; while in the domain
of $\left[ j-1/2,j\right] $, the energy level of $\left\vert
j,j\right\rangle $ is lower than both $ \left\vert
j,j-1\right\rangle $ and $\left\vert j,j-2\right\rangle $, and the
order of the energy levels is
$\omega_{j}<\omega_{j-1}<\omega_{j-2}<\cdots \cdots<\omega_{-j}$. We
also note that, as the coupling strength $g_{0}$ increases,
$g^{(2)}(\tau,0)$ oscillates faster with respect to $\delta$. In addition, as $N$
increases, the value of $g^{(2)}(\tau,0)$ increases.

Above, we gave a qualitative analysis of the generalized second-order coherence function based on perturbation theory. According to our calculations, there is a
large discrepancy between the theoretical analysis and the exact
numerical result. The reason may be as follows. As seen in the
definition of the generalized second-order coherence function, [i.e., Eq.~(\ref{g_2_t})], this is determined
by two correlation functions, that is, $ \langle \psi ^{\prime
}(0)|U_{I}^{c\dag }(\tau )J_{+}J_{-}U_{I}^{c}(\tau )|\psi ^{\prime
}(0)\rangle $\ and $\langle \psi (0)|U_{I}^{c\dag }(\tau
)J_{+}J_{-}U_{I}^{c}(\tau )|\psi (0)\rangle $. As far as the latter
is concerned, we calculate it in the interaction picture. Here, we
approximate the time-dependent wave function $U_{I}^{c}(\tau )|\psi
(0)\rangle $\ to first-order by perturbation
theory. Since the operator $ J_{+}J_{-}$ gives two large and
markedly different eigenvalues to the components $ |j,j\rangle $\
and $|j,j-1\rangle $, the originally small deviation in the
approximate wave function with respect to the exact one will be
enlarged.

However, when we come to the case with $m=0$ and $1$, the
situation turns out to be totally different. First of all, let us
turn to the Hamiltonian $ H=H_{0}+H_{I}$ given in Eqs.~(\ref{H0})
and~(\ref{H1}). In the large-detuning regime, it only induces a Rabi
oscillation between the two nearly degenerate states $|j,0\rangle $
and $|j,1\rangle $, while leaving the populations in the other
states almost unchanged. On account of the conservation of the total
probability and the same eigenvalues of the operator $J_{+}J_{-}$ on
the two relevant states, in the system which is initially in an
equal superposition of $|j,0\rangle $ and $|j,1\rangle $, the approximate correlation function $\langle \psi (0)|U_{I}^{\dag
}(\tau )J_{+}J_{-}U_{I}(\tau )|\psi (0)\rangle $ is expected to be
quite close to the exact one. This situation will not take place for
the case with $m=j$ and $j-1$, since the relevant eigenvalues of the
operator $J_{+}J_{-}$ are remarkably different from each other.
A similar analysis can be applied to the numerator in the generalized second-order coherence function. Consequently, the generalized second-order coherence function obtained from the perturbation theory
will coincide with the exact one for the case with $m=0$ and $1$,
while there is an obvious difference between the results from these
two methods for the case with $m=j$ and $j-1$. Therefore, we only give
the numerical results in Fig.~\ref{fig7}.

\section{\label{sec:6}Conclusion and remarks}

In this paper, we study the statistical properties of atomic
excitations for two cases: with dissipation and without dissipation. We
find that this statistical property can be controlled by the average photon number in
the cavity. In
addition, the photon-induced second-order interaction
between atoms is valid in the long range and can be strengthened by
a high-finesse microcavity with a very small effective
mode volume. % This may
%improve atomic excitations blockade induced by the dipole-dipole
%interaction\cite{Rydberg blockade1,Rydberg blockade2,Rydberg blockade3}.
Furthermore, we find that the double atomic excitation will be
suppressed when the average photon number in the cavity is in the vicinity of some
special points (degenerate points). We have also studied the
critical behavior of this statistical property of atomic excitations
around the critical point at which the QPT occurs.

To characterize the statistical property of atomic excitations, we
define a generalized second-order coherence function similar to the second-order coherence function for
photons. Furthermore, in the limit of $N\rightarrow\infty$ and low
excitations, it becomes the conventional one. We have demonstrated
the antibunching effect for atomic excitations near the degenerate
points and the characteristic of sub-Poissonian statistics, which
implies the existence of the atomic excitation blockade. Since this system possesses several critical points, we also study the critical behavior of
the generalized second-order coherence function of atomic excitations around the critical points. Our results
show the sensitivity of the system dynamics with the average photon number in the cavity.

\begin{acknowledgments}
We thank J.-N. Zhang and Chengyun Cai for helpful
discussions in numerical calculations, and M. Delanty and A. Miranowicz  for helpful discussions and very useful suggestions on the manuscript. The work is supported by the National Natural Science Foundation of China under Grants No.
10935010 and 11074261. F.N. acknowledges partial support from the Laboratory of Physical
Sciences, National Security Agency, Army Research Office, National Science
Foundation Grant No. 0726909, Grant-in-Aid for Scientific
Research (S), MEXT Kakenhi on Quantum Cybernetics, and the JSPS-FIRST program.
\end{acknowledgments}

\appendix*

\section{EXPLICIT EXPRESSIONS FOR THE PARAMETERS OF $g^{(2)}$}

In this appendix, we present the expressions for the parameters
used in Eqs.~(\ref{g2}) and~(\ref{g2QPT}), respectively.

For $m=0,1$, the parameters of $g^{(2)}(\tau,0)$ given by
Eq.~(\ref{g2}) are
%\begin{widetext}
\begin{eqnarray}
x_{1}&=&v_{0}^{2}(j-1)^{2}(j+2)^{2}\vert c_{1}\vert ^{2}\left\vert \eta_{1}\eta_{3}O_{2,0}
+\eta_{2}\eta_{4}O_{2,1} \right\vert ^{2},  \notag \\
x_{2}&=&v_{_{0}}^{2}(j+1) ^{2}j^{2}\left\vert c_{0}\right\vert
^{2}\left(
\frac{\eta_{3}\eta_{4}}{\eta_{2}\eta_{3}-\eta_{1}\eta_{4}}\right)
^{2}\left\vert O^{*}_{-1,0} -O^{*}_{-1,1}
\right\vert ^{2},  \notag \\
x_{3} &=&j^{2}(j+1) ^{2}\left\vert \frac{v_{0}c_{0}\left(
\eta_{4}\eta_{1}O^{*}_{-1,1}-\eta_{3}\eta_{2}O^{*}_{-1,0}\right)}{\eta_{2}\eta_{3}-\eta_{1} \eta_{4}}+c_{1}\right\vert ^{2},  \notag \\
x_{4} & =&j(j+2)(j^{2}-1)\left\vert v_{0}c_{1}\left(
\eta_{3}^{2}O_{-1,0}+\eta_{4}^{2}O_{-1,1}
\right)+c_{0}\right\vert ^{2},  \notag \\
x_{5} & =&v_{0}^{2}(j-1)(j^{2}-4)(j+3)\frac{\vert
c_{0}\vert^{2}}{\omega_{-2,-1}^{2}} \left\vert
1-e^{i\omega_{-2,-1}\tau}\right\vert ^{2},
\end{eqnarray}
and
\begin{eqnarray}
y_{1} & =&(j+1)j,  \notag \\
y_{2} & =&\Omega^{2}\left( j-1\right) ^{2}\left( j+2\right)
^{2}\nonumber\\&& \times\left\vert
\eta_{1}(c_{0}\eta_{3}+c_{1}\eta_{1})O_{2,0}
+\eta_{2}(c_{0}\eta_{4}+c_{1}\eta_{2})O_{2,1}
\right\vert ^{2},  \notag\\
y_{3} & =&v_{0}^{2}(j-1)(j+2)\nonumber\\&& \times\left\vert
\eta_{3}\left( c_{0}\eta_{3}+c_{1}\eta_{1}\right) O_{-1,0}
+\eta_{4}\left( c_{0}\eta_{4}+c_{1}\eta_{2}\right) O_{-1,1}
\right\vert ^{2},\nonumber\\
\end{eqnarray}
where
\begin{eqnarray}
v_{_{0}}\equiv\Omega\sqrt{(j+1)j},\hspace{0.5 cm}
O_{m,n}\equiv\frac{1}{\Delta_{m,n}}(1-e^{i\Delta_{m,n}\tau}).
\end{eqnarray}
%\end{widetext}
For $m=j-1,j$, the parameters of $g^{(2)}(\tau,0)$
given by Eq.~ (\ref{g2QPT}) are listed as follows:
\begin{eqnarray}
x_{1}^{c} & =&2j\left\vert a_{0}\left( \tau\right) \right\vert ^{2},  \notag
\\
x_{2}^{c} & =&2\left( 2j-1\right) \left\vert a_{1}\left( \tau\right)
\right\vert ^{2},  \notag \\
x_{3}^{c} & =&3\left( 2j-2\right) \left\vert a_{2}\left( \tau\right)
\right\vert ^{2},  \notag \\
x_{4}^{c} & =&4\left( 2j-3\right) \left\vert a_{3}\left( \tau\right)
\right\vert ^{2},  \notag \\
x_{5}^{c} & =&-2\alpha\left( \tau\right) \text{Im}\left[ a_{0}\left(
\tau\right) a_{1}^{\ast}\left( \tau\right) \right] ,  \notag \\
x_{6}^{c} & =&2\beta\left( \tau\right) \text{Re}\left[ a_{0}\left(
\tau\right) a_{1}^{\ast}\left( \tau\right) \right] ,  \notag \\
x_{7}^{c} & =&\gamma\left( \tau\right)[ \left\vert a_{0}\left(
\tau\right) \right\vert ^{2}-\left\vert a_{1}\left( \tau\right)
\right\vert ^{2}],
\end{eqnarray}
and
\begin{eqnarray}
y_{1}^{c}&=&j, \hspace{0.5 cm}
y_{2}^{c}=2j-1,  \notag \\
y_{3}^{c}&=&3\left( 2j-2\right) \left\vert c_{2}\right\vert ^{2},
\hspace{0.5 cm} y_{4}^{c}=\beta\left(\tau\right) ,
\end{eqnarray}
where
\begin{eqnarray}  \label{c2}
a_{0}(\tau)&=&\eta_{3}^{c}\eta_{4}^{c}f^{c}\left(O^{c*}_{j-2,0}-O^{c*}_{j-2,1}\right),  \notag \\
a_{1}(\tau) &=&\sqrt{j}-f^{c}\left(
\eta_{2}^{c}\eta_{3}^{c}O^{c*}_{j-2,0}
-\eta_{1}^{c}\eta_{4}^{c}O^{c*}_{j-2,1}\right),
\notag \\
a_{2}(\tau)&=&\sqrt{2j-1}\left[1+\frac{f^{c}\sqrt{j}}{2j-1} h_{1}(\tau)\right] ,  \notag \\
a_{3}(\tau) &=&\Omega\sqrt{3(2j-1)(2j-2)
}\frac{(1-e^{i\omega_{j-3,j-2}\tau})}{
\omega_{j-3,j-2}},  \notag \\
c_{2}(\tau) &=&\frac{f^{c}h_{2}(\tau) }{\sqrt{2(2j-1)}},
\end{eqnarray}
and
\begin{eqnarray}
\alpha(t) & =&q_{0}q^{-1} \sin(qt),  \notag \\
\beta(t) & =&q_{0} q ^{-2}(\omega_{j-1}-\omega_{j})\left[\cos(qt) -1\right],  \notag \\
\gamma(t) & =&2q_{0}q ^{-2}\Omega\sqrt{2j}\left[\cos(qt) -1\right],
\end{eqnarray}
with
\begin{eqnarray}
q_{0}&\equiv&-2\Omega\sqrt{2j}(j-1),\hspace{0.5 cm}
f^{c}\equiv\frac{\sqrt{2}\Omega(2j-1) }{
\eta_{2}^{c}\eta_{3}^{c}-\eta_{1}^{c}\eta_{4}^{c}}, \nonumber \\
q&\equiv&\sqrt{(\omega_{j-1}-\omega_{j})^{2}+8j\Omega^{2}},\hspace{0.3
cm}
O_{m,n}^{c}\equiv\frac{1-e^{i\Delta_{m,n}^{c}\tau}}{\Delta_{m,n}^{c}}.
\end{eqnarray}
Here,
\begin{eqnarray}
h_{1}(\tau)&=&\eta_{2}^{c}\eta_{3}^{c}O^{c}_{j-2,0} -\eta_{1}^{c}\eta_{4}^{c}O^{c}_{j-2,1},\notag \\
h_{2}(\tau)&=&\eta_{3}^{c}(\eta_{2}^{c}-\eta_{4}^{c})O^{c}_{j-2,0}
-\eta_{4}^{c}(\eta_{1}^{c}-\eta_{3}^{c})O^{c}_{j-2,1}.
\end{eqnarray}

%\end{CJK*}
\end{document}